\documentclass[aip, reprint]{revtex4-2}

\usepackage{amsmath, amsthm, bm}
\usepackage{amssymb}
\usepackage{mathrsfs}
\usepackage{amsfonts}
\usepackage{graphicx}
\usepackage[caption=false]{subfig}
\usepackage[dvipsnames]{xcolor}
\usepackage{siunitx}
\usepackage{bbold}
\usepackage[version=3]{mhchem}
\usepackage[acronym]{glossaries}
\usepackage{array}
\usepackage{blkarray}
\usepackage[section]{placeins}
\usepackage{indentfirst}
\usepackage{xspace}

\usepackage{xcolor}

\DeclareSIUnit\au{\text {au}}
\DeclareSIUnit\angstrom{\text {Å}}
\def\mgh{\ce{MgH+}}

\newacronym{bo}{BOA}{Born-Oppenheimer approximation}
\newacronym{cbo}{CBOA}{cavity Born-Oppenheimer approximation}
\newacronym{dse}{DSE}{dipole self-energy}
\newacronym{pes}{PES}{potential energy surface}
\newacronym{cpes}{cPES}{cavity potential energy surface}
\newacronym{vsc}{VSC}{vibrational-strong coupling}
\newacronym{esc}{ESC}{electronic-strong coupling}
\newacronym{lp}{LP}{lower polariton}
\newacronym{up}{UP}{upper polariton}
\newacronym{jc}{JC}{Jaynes-Cummings}
\newacronym{tc}{TC}{Tavis-Cummings}
\newacronym{fc}{FC}{Franck-Condon}
\newacronym{etc}{ETC}{extended Tavis-Cummings}
\newacronym{rwa}{RWA}{rotating wave approximation}
\newacronym{tls}{TLS}{two-level system}
\newacronym{cs}{CS}{coherent state}
\newacronym{cas}{CASSCF}{complete active space self-consistent field}
\newacronym{mrci}{MRCI}{multiconfiguration reference configuration interaction}
\newacronym{stirap}{STIRAP}{Stimulated Raman Adiabtic Passage}
\newcommand{\br}[1]{\langle #1 \vert}
\newcommand{\ke}[1]{\vert #1 \rangle}
\newcommand{\bk}[2]{\langle #1 \vert #2 \rangle}
\newcommand{\out}[2]{\vert #1 \rangle \langle #2 \vert }
\newcommand{\ev}[1]{\langle #1 \rangle}

\newcommand{\cm}[2]{\left[#1,#2\right]}
\DeclareMathOperator{\Tr}{Tr}

\newcommand{\expcv}[3]{\langle #1 \vert #2 \vert #3 \rangle}

\newcommand{\destroy}{\hat a}
\newcommand{\create}{\hat a^\dagger}

\newcommand{\sigmap}{\hat\sigma^+}
\newcommand{\sigmam}{\hat\sigma^-}

\DeclareUnicodeCharacter{0263}{\textgamma} %
\newcommand{\hamilt}{\hat{\mathcal H}}

\newcommand{\eV}{~\text{eV}}
\newcommand{\meV}{~\text{meV}\xspace}

\newcommand{\ps}{~\text{ps}\xspace}

\begin{document}

\title{
Selective excitation of molecular vibrations via a two-mode cavity Raman scheme
}

\author{Lucas Borges}
\affiliation{Department of Physics, Stockholm University, AlbaNova University Center, SE-106 91 Stockholm, Sweden}

\author{Thomas Schnappinger}
\email{thomas.schnappinger@fysik.su.se}
\affiliation{Department of Physics, Stockholm University, AlbaNova University Center, SE-106 91 Stockholm, Sweden}

\author{Markus Kowalewski}
\email{markus.kowalewski@fysik.su.se}
\affiliation{Department of Physics, Stockholm University, AlbaNova University Center, SE-106 91 Stockholm, Sweden}

\date{\today}%

\begin{abstract}
The experimental realization of strong light-matter coupling with molecules initiated the rapidly evolving field of molecular polaritonics.
Most studies focus on how exciton polaritons, which combine electronic excitations with confined light modes, alter photochemistry.
In this paper, we investigate their use in selectively exciting molecular vibrational states in the ground state.
Selectively exciting molecules to high vibrational states with infrared lasers to catalyze ground-state chemical reactions is a challenging task.
Here, we propose a two-cavity mode setup in the electronic strong coupling regime inspired by the process of Stimulated Raman Adiabatic Passage (STIRAP) to selectively populate excited vibrational states.
One cavity mode actively pumps the molecular system, while the other provides a highly effective and tunable decay channel via photon leakage.
We demonstrate the ability to selectively populate vibrational states for coherent and incoherent light sources using a molecular model system.
Our initial findings show high efficiency and suggest a possible route to steering and controlling chemical reactions in the electronic ground state based on electronic strong coupling.
\end{abstract}

\maketitle
\section{Introduction}
As synthetic pathways in modern chemistry become more complex~\cite{Trauner2018-bd}, the use of light to modify and control chemical reactions may offer a way to shortcut synthetic processes~\cite{Buglioni2022-em}.
Molecular vibrations and their dynamics define chemical reactivity, and controlling them with external light is a long-standing goal of photochemistry~\cite{Tannor1985-ow,Warren1993-qo,Rabitz2000-or,Ruetzel2011-zt}.
Although it is possible to trigger chemical reactions by selectively exciting reactants to high vibrational states with infrared (IR) light, as proposed in theoretical studies~\cite{Gollub2010-xj,Debnath2013-am,Thallmair2017-gi,Keefer2019-vt}, the experimental realization is rather challenging.
This is in part due to the presence of competing processes that transfer vibrational energy to other degrees of freedom, such as the internal modes of the molecule or the external modes of surrounding molecules~\cite{Owrutsky1994-rj}, and the need for sophisticated pulse-shaping techniques.
Consequently, only a few experimental observations of selective IR photochemistry have been reported~\cite{Witte2003-uk,Stensitzki2018-to,Delor2014-yz,Heyne2019-xr}.
For example, it was possible to accelerate a bimolecular ground state reaction that leads to the formation of a urethane derivative using a femtosecond IR laser pulse tuned to a carbonyl vibration~\cite{Stensitzki2018-to,Heyne2019-xr}.

In the last decade, a new way to control molecular properties using light has emerged: strong light-matter coupling between the vacuum electromagnetic field in an optical cavity and a molecular electronic transition, named \gls{esc}, or a vibrational transition, named \gls{vsc}, in a molecular ensemble~\cite{Herrera2020-bg,Garcia-Vidal2021-qe,Li2022-gi}.
This collective strong coupling leads to the formation of hybrid light-matter states known as molecular polaritons.
Under \gls{esc}, these polaritons can alter processes such as electron transport~\cite{Orgiu2015-fd,Eizner2019,Polak2020-ac,Wellnitz2021-xo,Sokolovskii2023-zk,Balasubrahmaniyam2023-nc,Wallner2024-qg,Koessler2025-pz}, light harvesting~\cite{Esteso2021-ox,Wu2022-xs}, or energy transport~\cite{Zhong2016-ao,Zhang2024-kt,Krupp2025-ex}.
For \gls{vsc}, experiments show that molecular properties~\cite{Fukushima2022-ox,Patrahau2024-xk} and chemical reactivity~\cite{Thomas2019-ve,Ahn2023-qk} can be modified.
Interestingly, the same ground-state reaction that could be accelerated using a femtosecond IR pulse~\cite{Stensitzki2018-to} was decelerated under \gls{vsc} conditions when the quantized light mode was coupled resonantly with a carbonyl vibration~\cite{Ahn2023-qk}.
This clearly indicates that fundamentally different mechanisms are at play, modifying the chemical reactivity.
As theoretically demonstrated by Li, Nitzan and Subotnik~\cite{Li2022-ds}, \gls{vsc} conditions can be utilized to selectively excite molecules to high vibrational states, which may offer the possibility of selective IR photochemistry.
Similarly, the formation of polaritons under \gls{vsc} can also provide a pathway to selectively populate IR-inactive modes~\cite{Ji2025-sw}.
Another proposed application of strong light-matter coupling that influences molecular reactivity is the use of a hybrid metallodielectric cavity setup~\cite{Ben-Asher2025-rx}.
In this setup, the interference of two cavity modes, which are constructed differently, gives rise to an energy-selective Purcell effect, which provides the ability to increase the yield of photoisomerization reactions.

In this manuscript, we follow the idea of using two nearly identical cavity modes to achieve highly selective excitation of specific vibrational states of a molecular system in the electronic ground state.
The principle used here is inspired by \gls{stirap} \cite{Gaubatz90jcp,Malinovsky1997-qn,Sola1999-dd,Hennrich00prl}: the two cavity modes couple to a common electronically excited state, making use
of quantum interference to suppress the population of the excited state.
The first cavity mode is resonant with the lowest possible transition (0-0 transition) between the ground state and the electronically excited state of the molecule, and it pumps the system.
The frequency of the second cavity mode is tuned to match the energy gap between the $v=0$ state of the electronic excited state and the desired vibrational state (e.g., $v=1$) of the electronic ground state.
This mode creates an efficient and selective relaxation channel via photon decay through the cavity mirrors for the excited system.

The paper is structured as follows.
After introducing the theoretical framework, we first demonstrate how the photon decay of a single cavity mode can selectively populate a vibrational state when the molecular system is driven by a continuous wave laser field.
Next, we replace the direct interaction between the molecule and the laser by a second cavity mode that is pumped either incoherently or coherently.
In the final part, we extend the analysis to a model with two molecules and demonstrate that it is possible to selectively excite collective vibrational states.

\section{Theory and models}

\begin{figure}
    \includegraphics[width=0.45\textwidth]{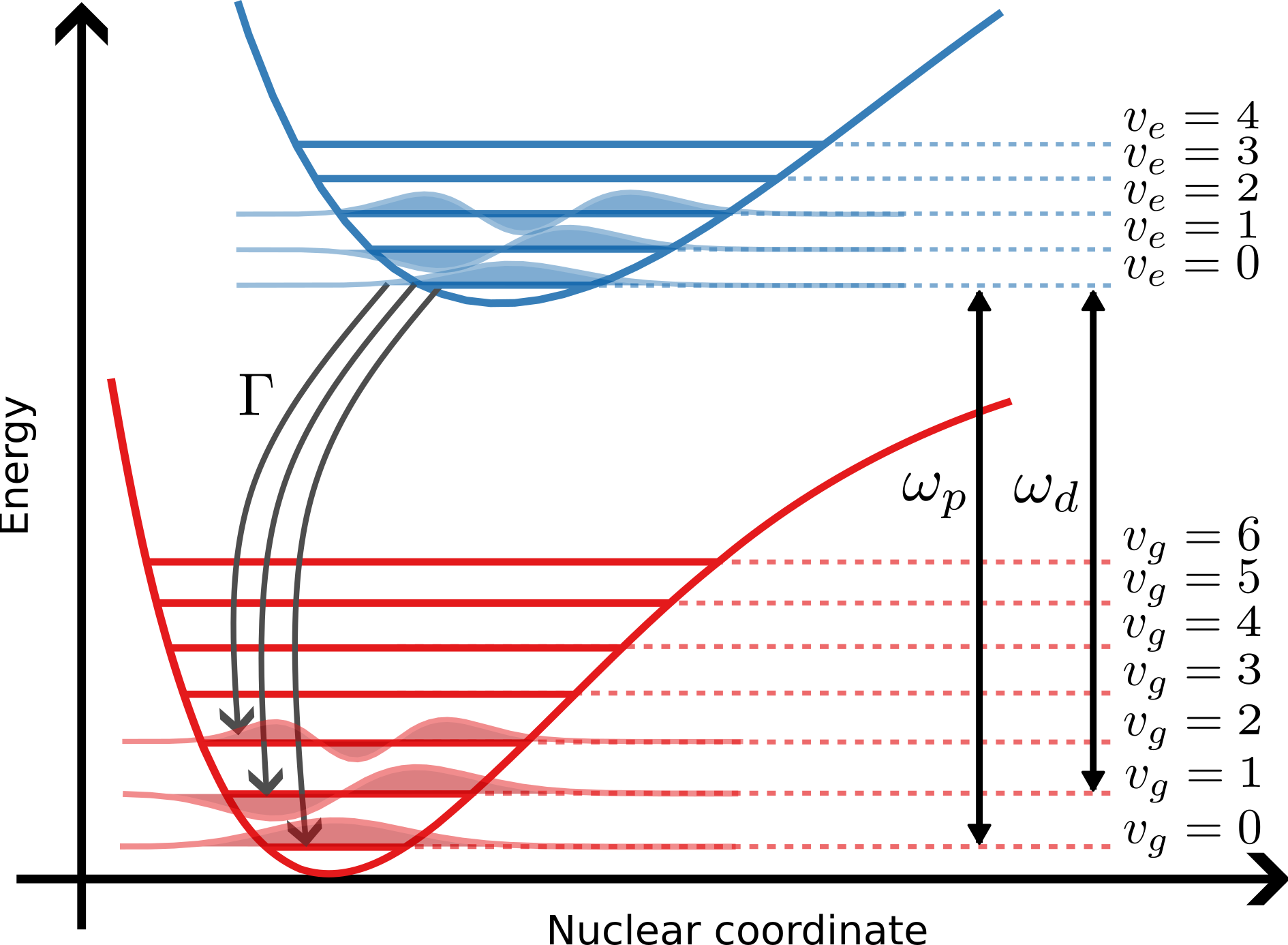}
    \caption{Energy diagram of the multilevel model formed by vibrational states of the first two electronic \glspl{pes} of \mgh molecule, $\ke{v_g}$ (red curve) and $\ke{v_e}$ (blue curve), coupled to two cavity modes indicated by the vertical arrows with frequency $\omega_p$ and $\omega_d$.
    The frequencies are chosen to match the energy difference between the first vibrational state in $\ke{e}$ and the
    first two vibrational states in $\ke{g}$: $\omega_p=4.32$\,eV and $\omega_d=4.12$\,eV, respectively (e.g., fundamental vibrational
    transition in $\ke{g}$ is 201\,meV).
    Possible decay channels via spontaneous emission are indicated by $\Gamma$.}
    \label{fig:ThreeLevelScheme}
\end{figure}

The molecular model in this work is based on the high-level ab initio \glspl{pes} of the \mgh molecule previously used to study the effect of \gls{esc} on molecular systems~\cite{Davidsson2020-qb,Davidsson2020-bs,Borges2024-pn}.
However, the reader should be aware that the molecular model used serves only as a proof of principle because it does not include potentially relevant factors such as vibrational relaxation.
We formulate the model consisting of two electronic states, $\ke{g}$ and $\ke{e}$, and their corresponding vibrational eigenstates.
This multilevel model and the underlying \glspl{pes} are schematically depicted in Fig.~\ref{fig:ThreeLevelScheme}.
In the following, we use atomic units ($\hbar = 4\pi \varepsilon_0 = m_e = 1$).
The uncoupled molecular Hamiltonian of the multilevel model is given by
\begin{equation} \label{eq:molecular_vibs}
\hamilt_M = \sum_{i=0}^{n_g-1}\omega_{g,i}\ke{v_g\!=\!i}\br{v_g\!=\!i}+ \sum_{j=0}^{n_e-1}\omega_{e,j}\ke{v_e\!=\!j}\br{v_e\!=\!j},
\end{equation}
where $\omega_{g,i}$ and $\omega_{e,j}$ are the eigenenergies of the molecular states and $n_g$ and $n_e$ are the number of vibrational states considered.
The states are labeled $\ke{v_g\!=\!i}$ and $\ke{v_e\!=\!j}$, for each electronic state, respectively.
We limit the number of vibrational states associated with each electronic level in the Hamiltonian by choosing the maximum values of $n_g$ and $n_e$ in Eq.~\ref{eq:molecular_vibs}.
For a single molecule, we use $n_g=7$ and $n_e=5$.
Figures~S5 and S6 in the Supporting Information show how the population dynamics change with an increasing number of vibrational states in the truncated Hamiltonian, and their change has converged
to $<10^{-3}$ for the chosen number of vibrational states.
When the model is extended to two molecules, we use $n_g=5$ and $n_e=3$ to reduce the computational time of propagation while maintaining sufficiently converged results.

The coupling of this multilevel model to two cavity modes is formulated using the \gls{jc} coupling scheme~\cite{Jaynes1963-re} in the dipole approximation and the rotating wave approximation.
A validation for the use of the rotating wave approximation in this work can be found in the Supporting Information.
As the two cavity modes used in this work have distinct tasks, we refer to the cavity mode used to drive the coupled system as pump ($p$) and the mode used to populate the vibrational states in $\ke{g}$ as dump ($d$).
The corresponding Hamiltonian reads
\begin{align}
    \label{eq:JCHamiltonian}
    \hamilt =&  \hamilt_M \\&+ \sum_{k\in\{p,d\}}\left[ \omega_k\create_k \destroy_k + \sum_{i=0}^{n_g-1}\sum_{j=0}^{n_e-1}g^{k}_{ij} \left(\create_k\sigmam_{ij} +\destroy_k\sigmap_{ij}\right)\right]. \nonumber
\end{align}
where the Pauli operators $\sigmam_{ij} =\out{v_g\!=\!i}{v_e\!=\!j}$ describe the transition between vibrational states in $\ke{g}$ and $\ke{e}$.
For each of these transitions, we define the coupling strength $g^{k}_{ij}$ as
\begin{equation} \label{eq:coupl_str}
    g^{k}_{ij} =\frac{ \mathcal{E}_c\mu_{ij}}{\sqrt{N}} = \sqrt{\frac{2\pi\omega_k}{NV_k}} \mu_{ij},
\end{equation}
with the frequency of the cavity mode $\omega_k$, its effective mode volume $V_k$, the number of molecules $N$ and $\mu_{ij} = \expcv{v_g\!=\!i}{\hat\mu_{eg}}{v_e\!=\!j}$ the Franck-Condon factor for the respective vibrational states (see Fig.~S1 in the Supporting Information).
For simplicity, we assume that the photon mode polarization and the transition dipole moment $\hat{\mu}_{eg}$ of the molecule are aligned parallel.
To quantify the coupling strength with a single scalar value, we use $g_{k} \equiv g^{k}_{00}$, which uses the Franck-Condon factor of the 0-0 transition for a given cavity frequency $\omega_k$.
The \gls{jc} coupling scheme can be extended beyond the single molecule case to include $N$ identical molecular systems, leading to the Tavis-Cummings coupling scheme~\cite{Tavis1967-op}. 

To include spontaneous decay of the electronic excited molecule and the photon decay from the cavity modes, we employ the Lindblad master equation to describe the dynamics of such an open system~\cite{Scala2007, Betzholz2020}
\begin{equation} \label{eq:master_equation}
\begin{split}
        \dot \rho(t) =& -i\cm{\hamilt(t)}{\rho(t)}\\ &+
    \sum_{k\in\{p,d\}}\frac{\kappa_k}{2}\left(\left[\create_k\destroy_k,\rho(t)\right]_+   - 2\destroy_k\rho(t)\create_k \right)
    \\&+ \sum_{i=0}^{n_g-1}\sum_{j=0}^{n_e-1}\frac{\Gamma_{ij}}{2}
    \left(\left[\rho(t), \out{v_e\!=\!j}{v_e\!=\!j}\right]_+ \right.\\
    &-  \left.2 \out{v_g\!=\!i}{v_e\!=\!j}\rho(t)\out{v_e\!=\!j}{v_g\!=\!i}\right)
    \,
\end{split}
\end{equation}
where $\kappa_k$ is the photon decay rate for the mode $k$, $\Gamma_{ij}$ is the spontaneous decay rate from state $\ke{v_e\!=\!j}$ to $\ke{v_g\!=\!i}$ and $[ A,B ]_+$ denotes the anticommutator $AB+BA$.
The photon lifetime is set to 500\,fs ($\kappa=2\text{ps}^{-1}$), which corresponds to a quality factor $Q=65$ for a cavity frequency of $4.31$\eV (molecular 0-0 transition) \cite{Fox2013}.
This quality factor aligns with Fábry-Perot cavities often used in polaritonic experiments~\cite{Mony18JPCC}, which typically exhibit $Q$ values of 100 and below.
The spontaneous decay rates are calculated using the corresponding Einstein coefficients
\begin{equation}
    \Gamma_{ij} = \frac{4 }{3}\left(\frac{\Delta\omega_{ij}}{c}\right)^3|\mu_{ij}|^2,
\end{equation}
with $\Delta\omega_{ij} = \omega_{e,j}-\omega_{g,i}$ and $c$ is the speed of light.
For the \mgh molecule, the spontaneous decay rate $\Gamma_{00}$ for the transition $\ke{v_e\!=\!0}\rightarrow \ke{v_g\!=\!0}$ is $1/40\ps^{-1}$.
We increased spontaneous decay rates by a factor of $10^3$,
to match the excited state lifetimes typically found in condensed matter systems.

In the following, we study three different scenarios to drive the system.
In the first scheme, the molecule is directly pumped by a continuous-wave laser field.
In the second scenario, the continuous wave laser field drives the cavity mode $p$.
As a third scheme, the cavity mode $p$ is pumped by incoherent light.
For the first scenario, a laser field is coupled directly to the molecular electronic transition:
\begin{gather}\label{eq:mol_pump}
   \hamilt_L(t) = \mathcal{E}_L  \cos(\omega_Lt) \sum_{i=0}^{n_g-1}  \sum_{j=0}^{n_e-1}\mu_{ij} \left(\sigmam_{ij} +\sigmap_{ij}\right),
\end{gather}
where $\mathcal{E}_L$ is the electric field amplitude and
$\omega_L$ is the center frequency of the driving laser.
The corresponding Rabi frequency is $\Omega_{ij}=\mathcal{E}_L\mu_{ij}$.
Note that a direct excitation of the molecule is straightforward,
but may not be easy to realize in thinly layered Fabry–Pérot cavities.
Thus, for the second scheme, the cavity mode $p$ is coherently pumped by a laser field, and the corresponding Hamiltonian is given by
\begin{equation}\label{eq:mode_pump}
    \hamilt_{cp}(t) = \zeta \left(\create_p + \destroy_p\right)\cos(\omega_L t),
\end{equation}
where $\zeta$ determines the pump strength and $\omega_L$ is equal to the cavity frequency $\omega_p$.
The two pumping schemes described by Eq.~\ref{eq:mol_pump} and Eq.~\ref{eq:mode_pump} can be added to the Hamiltonian $\hamilt(t)$ as part of the Lindblad master equation given in Eq.~\ref{eq:master_equation}.
For the third case, the cavity mode $p$ is pumped with incoherent light
which is described by the Liouvillian term,
\begin{equation}\label{eq:Lincohrent}
    \mathcal{L}_p = \frac{\eta}{2}\left(\left[\destroy_p\create_p, \rho(t)\right]_+
    - 2\create_p\rho(t)\destroy_p \right)\,
\end{equation}
which is added to the Lindblad master equation (Eq. \ref{eq:master_equation}).
Here $\eta$ is the photon pump rate of mode $p$.
The open system dynamics described by Eq.~\ref{eq:master_equation} is solved using the QuTiP package~\cite{qutip}, version 5.0.2, function mesolve, which is based on the SciPy zvode ODE solver.
All calculations were performed in a reproducible environment using the Nix package manager together with NixOS-QChem~\cite{Kowalewski2022-zt} (commit 319ff67).

\section{Results and Discussion}

To evaluate the performance of the cavity-mediated selective population of vibrational states, we determine the population dynamics of coupled molecular cavity systems. 
The time-dependent population $\mathcal{P}\ke{\phi;n}(t)$ of each eigenstate $\ke{\phi;n}$ of the coupled molecular photonic system is calculated as $\mathcal{P}\ke{\phi;n_p,n_d} = |\bk{\Psi(t)}{\phi;n_p,n_d}|^2,$ where $\ke{\Psi(t)}$ is the total state of the system, $\phi$ indicates the molecular states $\ke{v_e}$ and $\ke{v_g}$, and $n_p$ and $n_d$ are the photon numbers of mode $p$ and $d$.

\subsection{Exciting vibrational states in a single molecule}

\begin{figure*}
    \centering
    \includegraphics[width=0.9\linewidth]{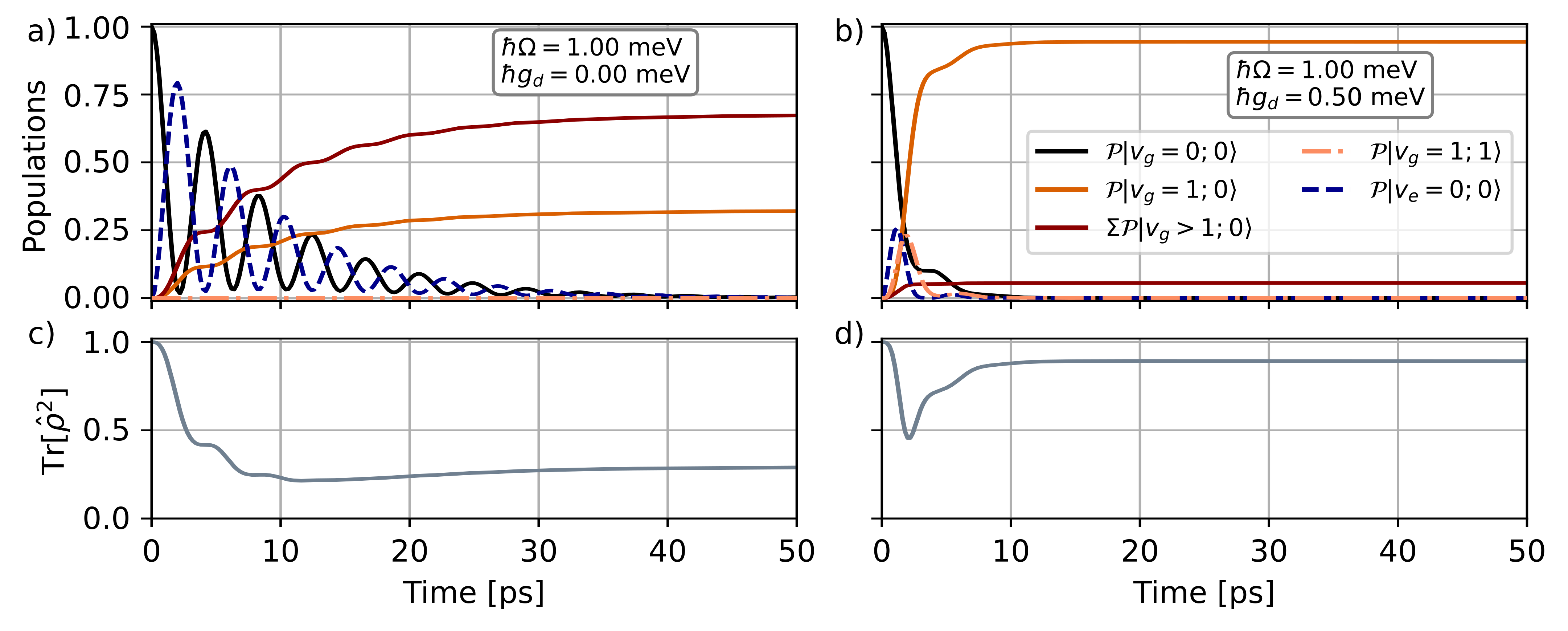}
    \caption{
    Time evolution of a single molecule directly driven by a laser field.
    Temporal evolution of the populations $\mathcal{P}\ke{\phi,n_d}$ (a)-(b) and the purity of the density matrix $\text{Tr}[\hat{\rho}^2]$ (c)-(d).
    The $\ke{v_g\!=\!0}\rightarrow\ke{v_e\!=\!0}$ transition is driven by a continuous wave laser field ($\omega_L=\Delta\omega_{00}$, $\Omega = \mathcal{E}_L\mu_{00}/\hbar = 1.00$\,meV) and the cavity mode $d$ is resonant with the $\ke{v_g\!=\!1}\rightarrow\ke{v_e\!=\!0}$ transition ($\omega_d = \Delta\omega_{10}$ and $\kappa_d=2.00$\,ps$^{-1}$).
    The case with $g_{d}=0.0$\,meV is shown in a) and c), and for $g_{d} = 0.5$\,meV in b) and d).
    The spontaneous decay rate $\Gamma_{00}$ is 1/40\,ps$^{-1}$. 
    }
    \label{fig:L1M_comparison_strength}
\end{figure*}

First, we investigate how a single cavity mode in the \gls{esc} regime, coupled to a single molecule, can be utilized to selectively populate vibrational states in the molecular ground state.
This cavity mode $d$ is chosen to be resonant with the transition $\ke{v_g\!=\!1}\rightarrow\ke{v_e\!=\!0}$ creating polaritonic/dressed states, and a continuous wave laser field directly drives the molecular transition $\ke{v_g\!=\!0}\rightarrow\ke{v_e\!=\!0}$ according to Eq.~\ref{eq:mol_pump}.
The results are given in Fig.~\ref{fig:L1M_comparison_strength} for two different
coupling strengths, $g_{d}=0.0$\,meV and $g_{d} = 0.5$\,meV.

Figure~\ref{fig:L1M_comparison_strength}(a) shows the population dynamics of the classical pumped molecular system without cavity coupling, i.e., $g_{d}=0.0$\,meV.  Rabi oscillations between $\ke{v_g\!=\!0; 0_d}\rightarrow\ke{v_e\!=\!0; 0_d}$ can be clearly observed, which are caused by the driving laser.
Here, the Rabi frequency of the driving laser is $\Omega_{00} = 1.00 \,\text{meV}/\hbar$, which corresponds to an intensity of $I = 1.63 \cdot 10^{8} \frac{\text{W}}{\text{cm}^2}$.
These oscillations are dampened by the spontaneous emission from $\ke{e}$, which leads to vibrational heating in $ \ke{g}$, caused by the nondiagonal Franck-Condon factors (see Fig.~S1 in the SI).
After 40\ps, approximately 30\% of the population is in $\ke{v_g\!=\!1; 0_d}$, while the remaining population is distributed in higher vibrational levels of the ground state $\ke{v_g\!>1; 0_d}$, and plotted as a dark red line in Fig.~\ref{fig:L1M_comparison_strength}(a).
The purity of the corresponding density matrix ($\Tr[{\rho^2}]$) in Fig.~\ref{fig:L1M_comparison_strength}(c) is decreasing and reaches a constant value of approximately $0.3$ after $t=40$\,ps, demonstrating that the state reached is a highly mixed state as expected for a vibrational hot state.
In Fig.~\ref{fig:L1M_comparison_strength}(b), the time evolution of the relevant populations with cavity coupling turned on ($g_{d} = 0.5$\,meV) is shown.
Resonantly coupling to the transition $\ke{v_e\!=\!0}\rightarrow\ke{v_g\!=\!1}$, the cavity mode $d$ provides a fast decay channel with a decay rate $\kappa_d$ of $2.00$\,ps$^{-1}$.
This cavity-mediated dissipation channel is two orders of magnitude faster than spontaneous emission.
Consequently, only a negligible part of the population is transferred to a state with a photon number $n_d$ greater than zero and only within the first 5\,ps a maximum of $\approx$25\% of the population is in $\ke{v_e\!=\!0; 0_d}$.
Already after 10\,ps $\approx$90\% of the population has been selectively transferred to $\ke{v_g\!=\!1; 0_d}$.
The heating effect is minimized and only 10\% of the population is transferred to higher vibrational states $\ke{v_g\!>1; 0_d}$.
The purity shown in Fig.~\ref{fig:L1M_comparison_strength}(d) confirms the high selectivity achieved, as it converges to approximately 0.9.

\begin{figure}
    \centering
    \includegraphics[width=.95\linewidth]{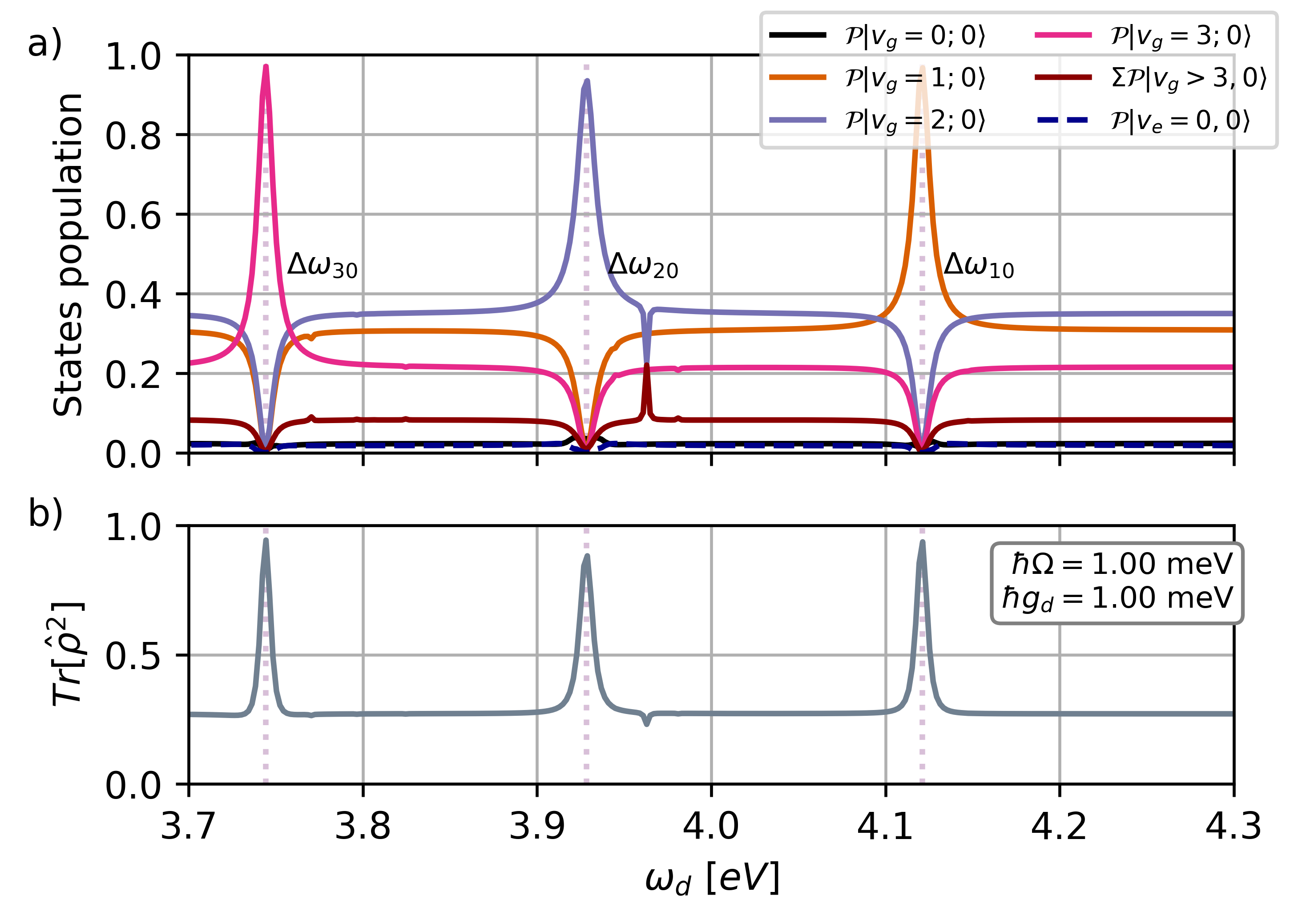}
    \caption{
    Frequency scan over the cavity mode $d$ for the directly pumped molecule.
    (a) Populations $\mathcal{P}\ke{\phi,n_d}$ and (b) purity of the density matrix $\text{Tr}[\hat{\rho}^2]$ as function of the cavity frequency $\omega_d$ after $t=30$\ps for a single molecule where the $\ke{v_g\!=\!0}\rightarrow\ke{v_e\!=\!0}$ transition is driven by a continuous wave laser field ($\omega_L=\Delta\omega_{00}$, $\Omega = \mathcal{E}_L\mu_{00}/\hbar = 1.00$meV) and coupled to the cavity mode $d$ ($g_d = 1.0$\,meV, $\kappa_d=2.00$\,ps$^{-1}$).
    The spontaneous decay rate $\Gamma_{00}$ is 1/40\,ps$^{-1}$, and the vertical dotted lines are indicating resonaces $\Delta\omega_{i0} = \omega_{e,0}-\omega_{g,i}$ with $i=1,2,3$.}
    \label{fig:L1M_scan-cavity}
\end{figure}

In addition to opening an effective decay channel, coupling to the cavity mode $d$ offers the possibility to select the final vibrational state by tuning the frequency $\omega_d$ of the cavity mode.
To demonstrate this, Fig.~\ref{fig:L1M_scan-cavity} shows the populations $\mathcal{P}\ke{\phi,n_d}$ and the purity of the density matrix $\text{Tr}[\hat{\rho}^2]$ after 30\,ps as a function of the cavity frequency $\omega_d$.

Sharp resonances are visible in both populations (see Fig.~\ref{fig:L1M_scan-cavity} (a)) and purity (see Fig.~\ref{fig:L1M_scan-cavity} (b)) when the cavity mode $d$ is resonant with a particular vibronic transition, i.e., $\ke{v_e\!=\!0}\rightarrow\ke{v_g\!=\!1...3}$, as indicated by the dotted lines.
At these resonances, more than 95\% of the population is transferred to the corresponding target state $\ke{v_g\!=\!1,2,3}$, and the purity of the density matrix exceeds 0.9.
In contrast, when the cavity is not resonant with any vibronic transition, the cavity mode has no visible influence.
Similarly to the results shown in Fig.~\ref{fig:L1M_comparison_strength}(a), the spontaneous emission takes over, resulting in a vibrationally hot state.
When the dump mode and pump laser are resonant with higher vibrational states in $\ke{g}$ and $\ke{e}$ (i.e., initial state with $v_g>1$), distinct, but weaker, peaks occur in the population for $\ke{v_g\!>3;\ 0_d}$  ($\omega_d=3.97$\,eV) that suddenly change the distribution of the vibrational state.

\begin{figure*}
    \centering
    \includegraphics[width=0.95\linewidth]{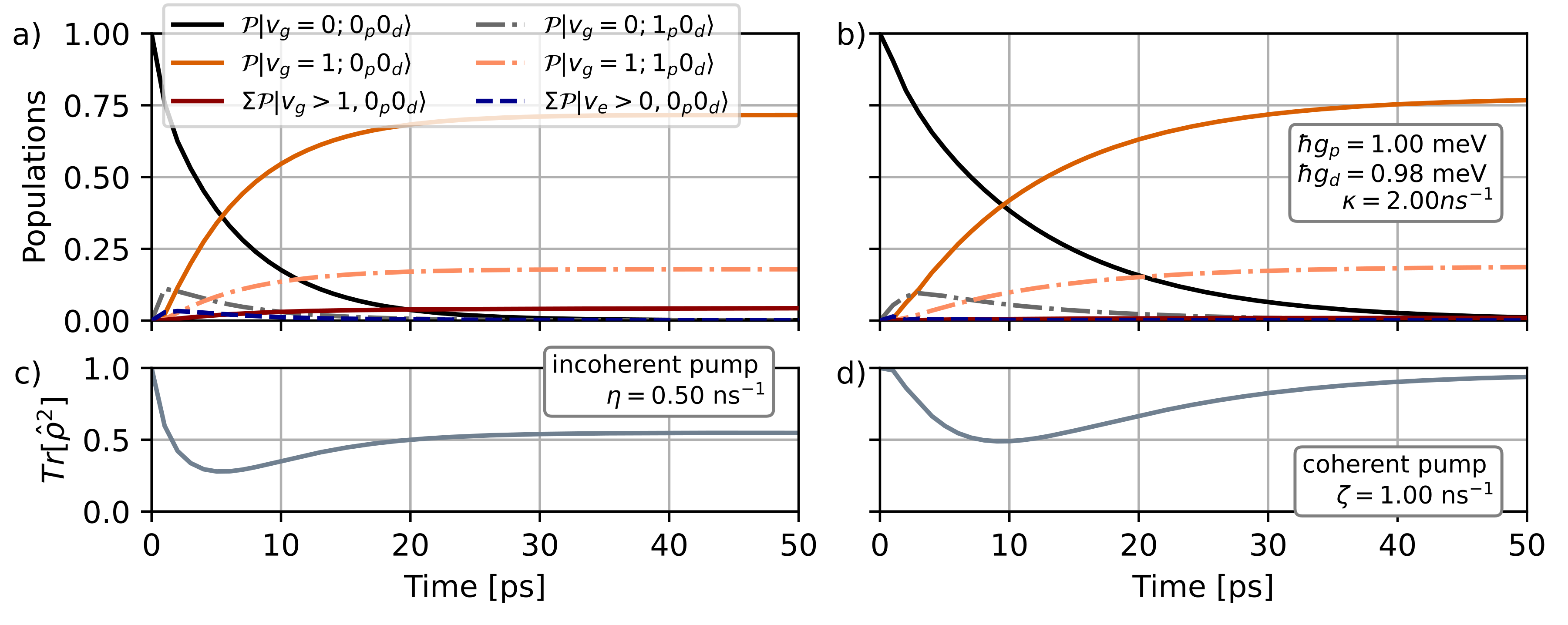}
    \caption{
    Time evolution of a single molecule coupled to two cavity modes.
    (a)-(b) Temporal evolution of the populations $\mathcal{P}\ke{\phi;n_p,n_d}$ and (c)-(d) the purity of the density matrix $\text{Tr}[\hat{\rho}^2]$.
    The two cavity modes $p$ and $d$ are tuned to $\omega_p=\Delta\omega_{00}$, $\omega_d=\Delta\omega_{10}$  and $\kappa_p = \kappa_d =2.00$\,ps$^{-1}$. For (a) and (c) the cavity mode $p$ is incoherently pumped according to Eq.~\ref{eq:Lincohrent} with $\eta=0.50$\,ps$^{-1}$. For b) and d), the cavity mode $p$ is coherently pumped according to Eq.~\ref{eq:mode_pump} with $\zeta=1.00$\,ps$^{-1}$.
    The spontaneous decay rate $\Gamma_{00}$ is 1/40 ps$^{-1}$.}
    \label{fig:TwoModes_comparison_pumps}
\end{figure*}

Next, we replace the classical laser field that directly pumps the molecule with a second cavity mode, $p$, which drives the coupled molecular system while leaving the second cavity mode, $d$, unpumped.
To investigate this two-cavity-mode setup, we set the cavity mode $p$ to be resonant with the $\ke{v_g\!=\!0}\rightarrow\ke{v_e\!=\!0}$ transition.
We compare a coherent pumping of the cavity mode $p$ (Eq.~\ref{eq:mode_pump}) to an incoherent pumping scheme (Eq.~\ref{eq:Lincohrent}).
Similarly to the results shown in Fig.~\ref{fig:L1M_comparison_strength} the other cavity mode $d$ is resonant with the $\ke{v_g\!=\!1}\rightarrow\ke{v_e\!=\!0}$ transition.
The coherent pump rate $\zeta$ of $1.00$\,ps$^{-1}$ and the incoherent photon pump rate $\eta$ of $0.50$\,ps$^{-1}$ are chosen to achieve a comparable photon number in cavity mode $p$.
Using the vacuum field strength $|\mathcal E_c| = \frac{g_p}{\mu_{00}} $ with $ g_p = 1 \, \text{meV} $, we find $|\mathcal E_c| = 3.5 \times 10^{7} \, \text{V/m} $.
The intra-cavity intensity of the empty cavity is given by
$ I \approx c\epsilon_0 |\mathcal E_c|^2 \left(\ev{\hat{n}}+0.5\right)$.
From the dynamics of an empty cavity subjected to photon pumping and decay, shown in Fig.~S8 of the Supporting Information, we determine the final number of photons to be $ \langle \hat{n} \rangle_\zeta = 0.250 $ for the coherent pumping scheme and $ \langle \hat{n} \rangle_\eta = 0.328 $ for the incoherent pumping scheme.
Consequently, the intra-cavity intensities created by the pumping are $ I_\zeta = 2.440 \times 10^{8} \, \text{W/cm}^2$ and $I_\eta = 2.694 \times 10^{8} \, \text{W/cm}^2$.
Due to the interplay of photon pumping, described by Eq.~\ref{eq:mode_pump} or Eq.~\ref{eq:Lincohrent}, and photon leakage, characterized by Lindbladian jump operators $\sqrt{\kappa_p} \destroy_p$ and $\sqrt{\kappa_d} \destroy_d$, an empty cavity reaches a steady-state photon number.
Based on the results of an empty cavity, we can truncate the Hamiltonian, considering only states with a maximum of two photons in mode $p$ and one photon in mode $d$, as these are the relevant states in the dynamics.
Note that the coupled system, strictly speaking, has no steady state, since we neglect the spontaneous decay of the vibrational states in $\ke{g}$.
The off-resonant heating that may occur due to this approximation happens on a much longer timescale.
The states reached within 100\,ps can thus be considered a good approximation of a steady state.
Fig.~\ref{fig:TwoModes_comparison_pumps} illustrates the evolution of the populations and the purity of the density matrix over time for both the incoherent pump scheme and the coherent pump scheme, employing the two-cavity-mode setup coupled to a single molecule.

The population dynamics for incoherent and coherent pumping, shown in Fig.~\ref{fig:TwoModes_comparison_pumps}(a) and (b) are qualitatively similar for the chosen cavity coupling strength.
In both cases, the ground state of the system $\ke{v_g\!=\!0; 0_p,0_d}$ is depopulated, and the target state $\ke{v_g\!=\!1;0_p,0_d}$  and its one-photon counterpart $\ke{v_g\!=\!1;1_p,0_d}$ become rapidly populated.
For an incoherent pump, 4\% of the population occupies states formed by higher vibrational states ($\ke{v_g\!>\!1;0_p,0_d}$).
Meanwhile, about 95\% of the population is distributed over the two states $\ke{v_g\!=\!1;0_p,0_d}$ and $\ke{v_g\!=\!1;1_p,0_d}$.
For the coherent pump, about 98\% of the population reaches these two states.
Although the population dynamics appear similar for both coherent and incoherent pumping, the differences are clearly visible when comparing the purity of the two density matrices $\Tr[\rho^2]$, shown in Figs.~\ref{fig:TwoModes_comparison_pumps}(c) and d).
For incoherent pumping, the system converges to a state with low purity ($\Tr[\rho^2] = 0.55$) while for coherent pumping, it converges to a value close to 1.

\begin{figure*}
    \centering
    \includegraphics[width=0.95\textwidth]{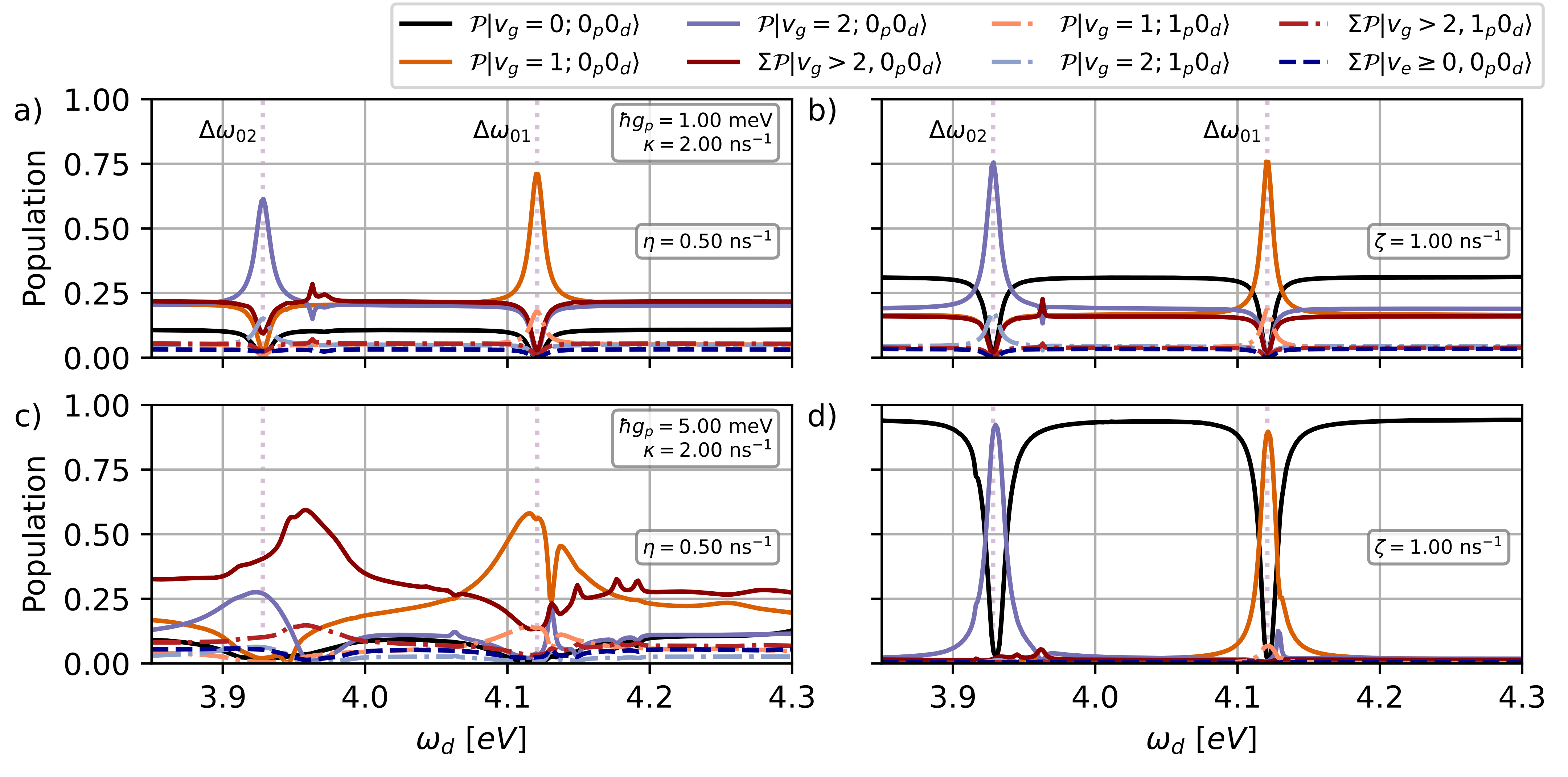}
    \caption{
    Frequency scan of cavity mode $d$ for a single molecule driven by cavity mode $p$.
    Populations $\mathcal{P}\ke{\phi,n_p,n_d}$ as function of the cavity frequency $\omega_d$ after $t=50$\,ps for a single molecule coupled the two cavity modes $p$ and $d$ with $\omega_p=\Delta\omega_{00}$, and $\kappa_p = \kappa_d =2.00$\,ps$^{-1}$. For a) and (c) the cavity mode $p$ is incoherently pumped according to Eq.~\ref{eq:Lincohrent} with $\eta=0.50$\,ps$^{-1}$. For b) and d), the cavity mode $p$ is coherently pumped according to Eq.~\ref{eq:mode_pump} with $\zeta=1.00$\,ps$^{-1}$.
    The coupling strengths of both modes are given by $g_d = \sqrt{\omega_d/\omega_p}\ g_{p}$, where $g_{p}$ is set to (a)-(b) $1.00$\,meV and (c)-(d) $5.00$\,meV. 
    The spontaneous decay rate $\Gamma_{00}$ is 1/40\,ps$^{-1}$, and the vertical dotted lines are indicating resonaces $\Delta\omega_{i0} = \omega_{e,0}-\omega_{g,i}$ with $i=1,2$.
    }
    \label{fig:TwoModes_Scan-w2}
\end{figure*}

To evaluate the selectivity of the proposed pump-dump scheme using two cavity modes, we scan the frequency of the cavity mode $d$ while keeping the cavity mode $p$ resonant with the $\ke{v_g\!=\!0}\rightarrow\ke{v_e\!=\!0}$ transition.
In Fig.~\ref{fig:TwoModes_Scan-w2} the populations $\mathcal{P}\ke{\phi;n_p,n_d}$ after 50\,ps as a function of the cavity frequency $\omega_d$ are shown.
The corresponding purity of the density matrix $\text{Tr}[\hat{\rho}^2]$ is given in the Supporting Information Fig.~S9.
Since the coupling strength depends on the cavity frequency (see Eq.~\ref{eq:coupl_str}), we rescaled the coupling strength $g_d$ relative to $g_p$ by a factor of $\sqrt{\omega_d/\omega_p}$ to achieve a comparable vacuum field strength for both modes while varying $\omega_d$.
We show two coupling strengths: $ g_p = 1.00$\,meV and of $g_p = 5.00$\,meV.
The first coupling strength is at the boundary between weak and strong coupling with a ratio of $2g_p/\kappa_p = 1.5$. The second coupling strength is clearly
in the strong coupling regime with $2g_p/\kappa_p = 7.6$
For both coherent pumping and incoherent pumping, the same pump rates of $\zeta=1.00$\,ps$^{-1}$ and $\eta=0.50$\,ps$^{-1}$ as in Fig.~\ref{fig:TwoModes_comparison_pumps} are used.

Figs.~\ref{fig:TwoModes_Scan-w2}(a) and (b) show the populations after 50\,ps for the incoherent pumping scheme ($\eta=0.50\ps^{-1}$) and the coherent pumping scheme ($\zeta=1.00\ps^{-1}$), respectively, with a cavity coupling strength of $g_{p} = 1.00$\,meV.
For this coupling, both schemes perform similarly and show an efficient population transfer to the selected vibrational state (i.e., $\ke{v_g\!=\!1,2;0_p,0_d}$), when the cavity mode $d$ is in resonance with either $\ke{v_g=0}$ or $\ke{v_g=1}$.
Coherent pumping results in a slightly higher population in the target states by suppressing vibrational heating more efficiently.
Compared to the direct molecular pump in Fig.~\ref{fig:L1M_scan-cavity}, the population in the desired vibrational target states is slightly reduced.
When the cavity mode $d$ is off-resonant with any vibronic transition, vibrational heating occurs in $\ke{g}$, as already observed in the single cavity mode setup.
However, for coherent pumping, the state with the highest population is $\ke{v_g\!=\!0;0_p,0_d}$, indicating less heating, even for an off-resonant cavity mode $d$.

When increasing the coupling strength to $g_p = 5.00\meV$, the difference between
the incoherent pump Fig.~\ref{fig:TwoModes_Scan-w2}(c) and the coherent pump Fig.~\ref{fig:TwoModes_Scan-w2}(d) are more pronounced.
For the incoherent pump, the population of higher vibrational states (dark red line) increases almost independently of the cavity frequency $\omega_d$.
Simultaneously, the resonances for specific vibronic transitions are much broader for the incoherent pump, and the overall efficiency of reaching a particular vibrational state is now reduced on resonance.
For the coherent pump shown in Fig.~\ref{fig:TwoModes_Scan-w2}(d), increasing the coupling strength of the cavity results in a more selective population of specific vibrational states $\ke{v_g\!=\!1,2;0_p,0_d}$ when $\omega_d$ is resonant with
the respective transition.
When the cavity mode $d$ is not in resonance, the stronger cavity coupling prevents vibrational heating, and approximately 90\% of the population remains in the $\ke{v_g\!=\!0;0_p,0_d}$ state.

The mechanism that drives the population transfer can be explained as follows.
A two-mode upper polariton $\ke{U}$ and a lower polariton state $\ke{L}$ are formed.
These two states are superpositions of $\ke{v_e=0;0_p,0_d}$, $\ke{v_g=0;1_p,0_d}$, and $\ke{v_g=1;0_p,1_d}$.
Moreover, a third eigenstate $\ke{M}$ is formed, which is a negative superposition of $\ke{v_g=0;1_p,0_d}$ and $\ke{v_g=1;0_p,1_d}$.
The state $\ke{M}$ has no contribution from the electronically excited and is responsible for efficient population transfer between the two vibrational states.
Note that this is analogous to the three-level scheme used in \gls{stirap}.
A visualization of the three polaritonic eigenstates can be found in the Supporting Information Fig.~S3.
The photon decay $\kappa_d$ results in a rapid decrease from $\ke{v_g=1;0_p,1_d}$ to $\ke{v_g=1;0_p,0_d}$, while $\ke{v_g=0;1_p,0_d}$ is actively pumped.
The quantum interference of the two cavity modes together with the photon decay is thus responsible for the directed population transfer.
The difference in selectivity between the incoherent and coherent pumping schemes is due to the selectivity to populate and depopulate the eigenstates $\ke{U}$, $\ke{L}$, and $\ke{M}$.
In the coherent pumping scheme, only $\ke{M}$ is populated, which leads to a selective decay via photon loss to the traget state.
In the incoherent pumping scheme $\ke{U}$ and $\ke{L}$ can also be populated.
Since these have contributions from the excited state $\ke{e}$, vibrational heating occurs, and the overall selectivity is reduced compared to the coherent pumping scheme (see Fig.~S4 in the Supporting Information).

\subsection{Exciting collective vibrational states in two molecules}

\begin{figure}
    \centering
    \includegraphics[width=\linewidth]{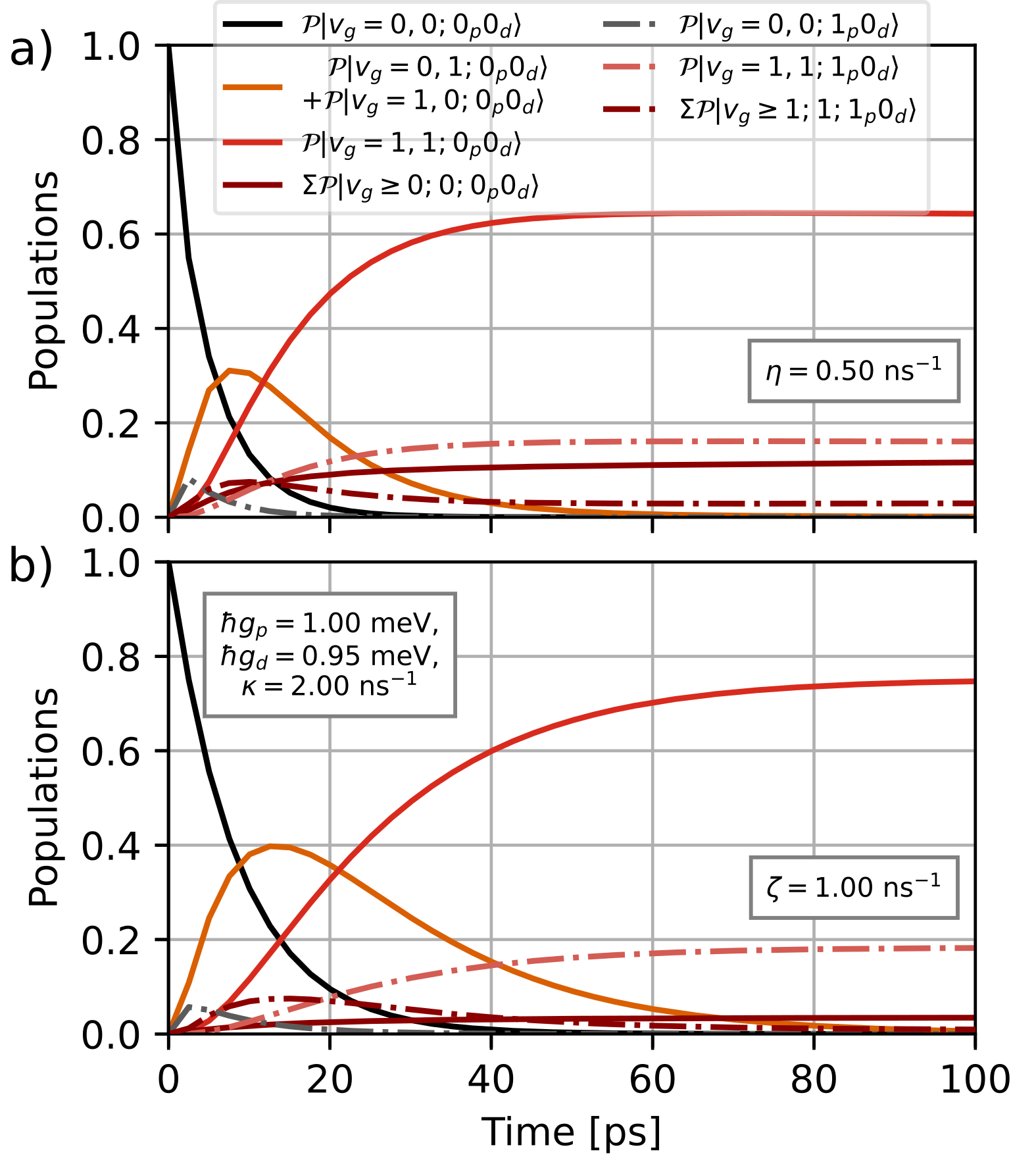}
    \caption{Temporal evolution of the populations of two molecules coupled the two cavity modes $p$ and $d$.
    The molecules coupled the two cavity modes $p$ and $d$ with $\omega_p=\Delta\omega_{00}$, $\omega_d=\Delta\omega_{10}$  and $\kappa_p = \kappa_d =2.00$\,ps$^{-1}$.
    The coupling strengths of both modes are given by
    $g_d = \sqrt{\omega_d/\omega_p}\ g_{p}$, with $g_{p}=1.00$\,meV.
    For (a) the cavity mode $p$ is incoherently pumped according to Eq.~\ref{eq:Lincohrent} with $\eta=0.50$\,ps$^{-1}$ and for (b) the cavity mode $p$ is coherently pumped according to Eq.~\ref{eq:mode_pump} with $\zeta=1.00$\,ps$^{-1}$.
    The spontaneous decay rate $\Gamma_{00}$ is 1/40 ps$^{-1}$.}
    \label{fig:TwoMol_TwoModes-comparison_psi0_5meV}
\end{figure}

In the final section of the manuscript, we expand our model system to include a second identical molecule.
The Hamiltonian from Eq. \ref{eq:JCHamiltonian} is extended to the Tavis-Cummings coupling scheme \cite{Tavis1967-op} and both molecules interact with the same two cavity modes, which act as pump and dump channels.
The only possible interaction between the two molecules included in our model occurs via the cavity modes.
Each molecule is described by five vibrational states in $\ke{g}$ and three vibrational states in $\ke{e}$.
The maximum number of photons is truncated to two in the cavity mode $p$ and one in the cavity mode $d$.
The corresponding total wavefunction describing the ground state of the coupled molecular cavity system reads $\ke{v_g\!=0,v_g\!=\!0;0_p,0_d}$.
Due to the symmetry of the molecular model, most collective states are formed by linear combinations, such as $\ke{v_g\!=1,v_g\!=\!0;0_p,0_d} \pm \ke{v_g\!=0,v_g\!=\!1;0_p,0_d}$, consisting of local, i.e., single-molecule excitations.
Since only one of the two possible linear combinations has a non-vanishing dipole moment and can interact with the cavity field, we define this one as the symmetric linear combination and restrict our discussion to these optically active states.
Similarly to the single-molecule case, we investigate the performance for incoherent and coherent pumping of cavity mode $p$ using the same pumping and decay rates as before.
The cavity mode $p$ is in resonance with the transition $\ke{v_g\!=\!0}\rightarrow\ke{v_e\!=\!0}$ and the cavity mode $d$ is in resonance with the transition $\ke{v_g\!=\!1}\rightarrow\ke{v_e\!=\!0}$.
The population dynamics obtained with incoherent and coherent pumping and a cavity coupling strength of $g_{p}=1.0$\,meV is given in Fig.~\ref{fig:TwoMol_TwoModes-comparison_psi0_5meV}.

\begin{figure}
    \includegraphics[width=0.75\linewidth]{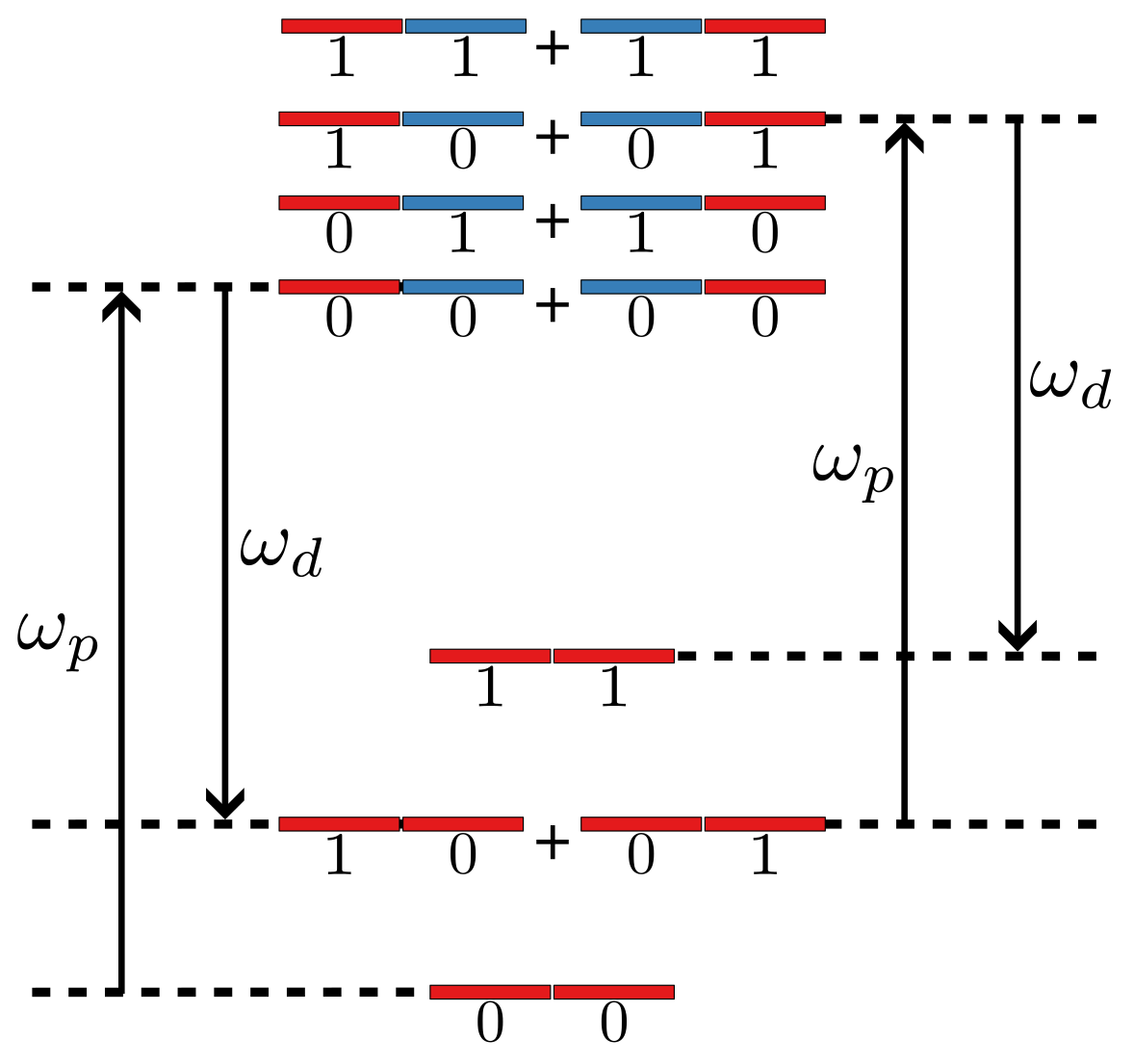}
    \caption{Diagram of the relevant bare (i.e., field-free) vibronic states of the multilevel model describing two identical molecules. The electronic states of each molecule are color-coded (red for $\ke{g}$ and blue for $\ke{e}$), and the numbers indicate the corresponding vibrational state of each molecule. Collective states formed by symmetric linear combinations are denoted by a plus sign.
    The possible interactions with the pump and dump cavity modes are labeled by the vertical arrows with frequency $\omega_p$ and $\omega_d$.}
    \label{fig:doublepump}
\end{figure}

Figs.~\ref{fig:TwoMol_TwoModes-comparison_psi0_5meV}(a) and (b) show the population in the first 100\,ps of the relevant collective states for coherent and incoherent pumping.
In both cases, initially the ground state $\ke{v_g\!=0,v_g\!=\!0;0_p,0_d}$ is quickly depopulated and the collective state  $\ke{v_g\!=1,v_g\!=\!0;0_p,0_d}\!+\! \ke{v_g\!=0,v_g\!=\!1;0_p,0_d}$ gains population.
However, this linear combination is only an intermediate, and most of the population is transferred to the state $\ke{v_g\!=1,v_g\!=\!1;0_p,0_d}$, for which both molecules are in the first excited vibrational state of $\ke{g}$.
For the case of coherent pumping (Fig.~\ref{fig:TwoMol_TwoModes-comparison_psi0_5meV}(b)) the population in the two polaritonic states formed by $\ke{v_g\!=1,v_g\!=\!1}$ saturates, and after 100\,ps it is larger than 95\%.
If the system is pumped incoherently,
the population of the two $\ke{v_g\!=1,v_g\!=\!1}$ states is approximately 80\% after 100\,ps and around 20\% of the population is transferred to higher vibrational states.
To understand the unexpectedly high population of states formed by $\ke{v_g\!=1,v_g\!=\!1}$ for coherent and incoherent pumping, we examine the structure of the bare, i.e., not hybridized, vibronic level in the case of two molecules.
Fig.~\ref{fig:doublepump} shows a schematic representation of the relevant states.

Starting in the collective ground state, $\ke{v_g\!=0,v_g\!=\!0}$ the coupling with the pump cavity mode $p$ excites the system into hybrid states, formed by the state $\ke{v_g\!=0,v_e\!=\!0}\!+\!\ke{v_e\!=0,v_g\!=\!0}$ as the mode $p$ is resonant with the $\ke{v_g\!=\!0}\rightarrow\ke{v_e\!=\!0}$ transition.
The second cavity mode $d$ tuned to the $\ke{v_g\!=\!1}\rightarrow\ke{v_e\!=\!0}$ transition opens an effective decay channel that leads to the population of the collective state $\ke{v_g\!=1,v_g\!=\!0}\!+\!\ke{v_g\!=0,v_g\!=\!1}$.
This pathway is illustrated on the left side of Fig.~\ref{fig:doublepump} and is identical to the one discussed and observed in the case of a single molecule.
However, since the cavity mode $p$ is also resonant with the transition from $\ke{v_g\!=1,v_g\!=\!0}\!+\!\ke{v_g\!=0,v_g\!=\!1}$ to $\ke{v_g\!=1,v_e\!=\!0}\!+\!\ke{v_e\!=0,v_g\!=\!1}$ and due to continuous pumping, the system gets excited a second time.
As the cavity mode $d$ also resonates with the transition from $\ke{v_g\!=1,v_e\!=\!0}\!+\!\ke{v_e\!=0,v_g\!=\!1}$ to $\ke{v_g\!=1,v_g\!=\!1}$, the decay of a second photon allows to populate the state $\ke{v_g\!=1,v_g\!=\!1}$.
This second cycle of pumping and dumping mediated by the same two cavity modes is illustrated on the right side of Fig.~\ref{fig:doublepump} and explains the high population of the state $\ke{v_g\!=1,v_g\!=\!1;0_p,0_d}$ for coherent and incoherent pumping.

\begin{figure*}
    \centering
    \includegraphics[width=0.95\textwidth]{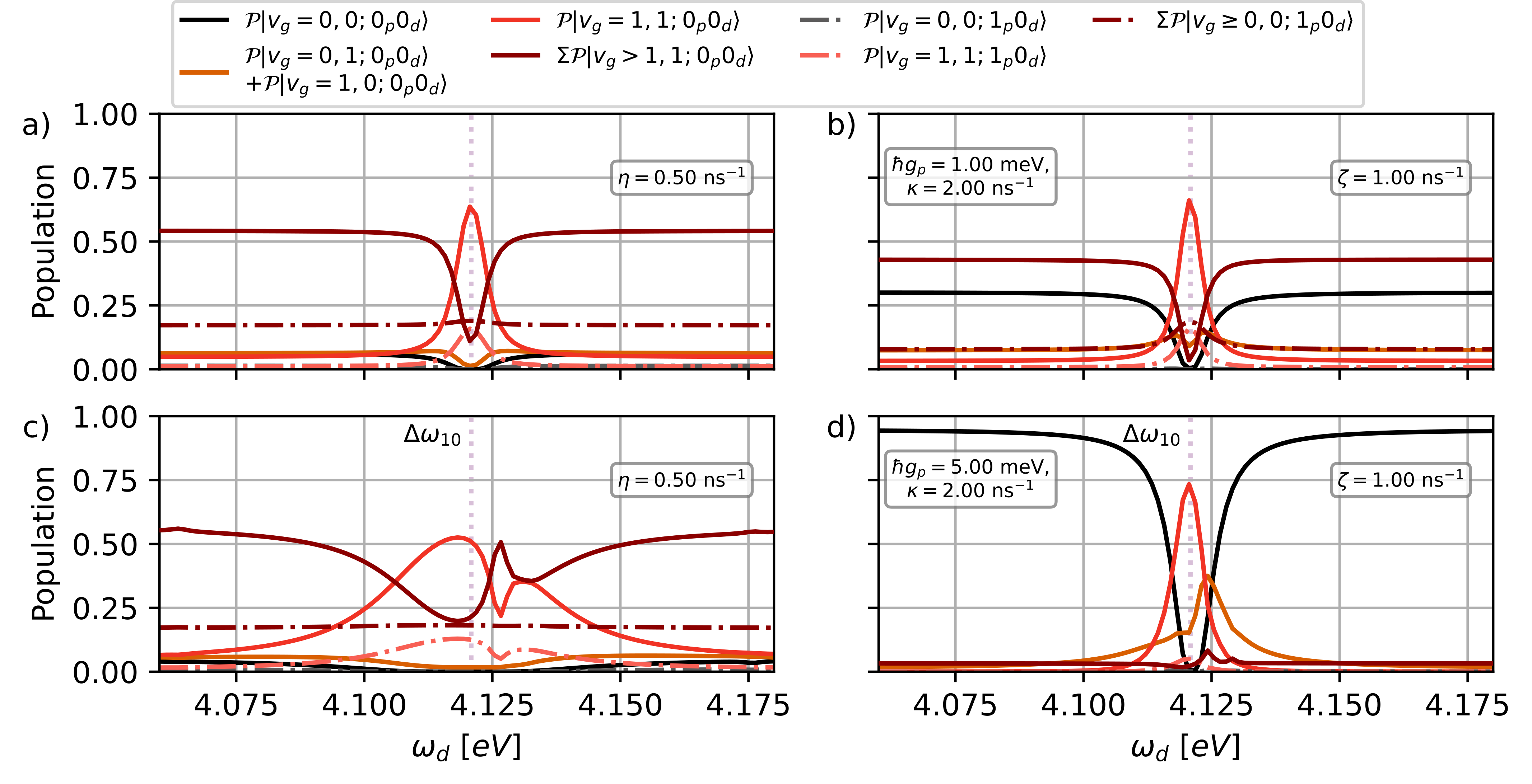}
    \caption{
Frequency scan of
Populations $\mathcal{P}\ke{\phi,n_p,n_d}$ as function of the cavity frequency $\omega_d$ after $t=50$\ps for two molecules coupled the two cavity modes $p$ and $d$ with $\omega_p=\Delta\omega_{00}$, and $\kappa_p = \kappa_d =2.00$\,ps$^{-1}$. For a) and c) the cavity mode $p$ is incoherently pumped according to Eq.~\ref{eq:Lincohrent} with $\eta=0.50$\,ps$^{-1}$. For b) and d), the cavity mode $p$ is coherently pumped according to Eq.~\ref{eq:mode_pump} with $\zeta=1.00$\,ps$^{-1}$.
The coupling strengths of both modes are given by $g_d = \sqrt{\omega_d/\omega_p}\ g_{p}$, where $g_{p}$ is set to a)-b) $1.00$meV and c)-d) $5.00$meV.
The spontaneous decay rate $\Gamma_{00}$ is 1/40 ps$^{-1}$, and the vertical dotted lines are indicating resonaces $\Delta\omega_{10} = \omega_{e,0}-\omega_{g,1}$.}
    \label{fig:TwoMol_scanw2_pumps-w2}
\end{figure*}

The effect of detuning the cavity $\omega_d$ from the vibronic transition $\ke{v_g\!=\!1}\rightarrow\ke{v_e\!=\!0}$ on the population after $t=50$\ps is shown in Fig.~\ref{fig:TwoMol_scanw2_pumps-w2}.
Similarly to the single molecule case, two coupling strengths, $g_p = 1.00$,meV and $5.00$,meV, are used, as well as both pumping schemes with a coherent pumping rate of $\zeta=1.00$\,ps$^{-1}$ and an incoherent pumping rate of $\eta=0.50$\,ps$^{-1}$.

The populations after $t=50$\ps are shown in Figs.~\ref{fig:TwoMol_scanw2_pumps-w2}(a) and (b) for $g_{p} = 1.00$\,meV.
Both pumping schemes perform similarly and show efficient population transfer to the vibrational state $\ke{v_g\!=1,v_g\!=\!1}$, when the cavity mode $d$ is in resonance with the vibronic transition $\ke{v_g\!=\!1}\rightarrow\ke{v_e\!=\!0}$.
However, in comparison to the single-molecule case, this process is less efficient.
In line with the single-molecule results shown in Fig.~\ref{fig:TwoModes_Scan-w2}, coherent pumping leads to a slightly higher population in the target states by suppressing vibrational heating more efficiently.
When the cavity mode $d$ is detuned, vibrational heating occurs in $\ke{g}$.
In the coherent pumping case, more population remains in the vibrational ground state than in the incoherent pumping case.
For the increased coupling strength of $g_p = 5.00\meV$, the results shown in Figs.~\ref{fig:TwoMol_scanw2_pumps-w2}c) and d) are very similar to those of a single molecule.
The performance of the two-cavity setup improves with larger coupling strengths when the cavity mode $p$ is coherently pumped, thus increasing population transfer on resonance and reducing vibrational heating.
For incoherent pumping, the population of higher vibrational states (dark red line) increases almost independently of the cavity frequency $\omega_d$.
The resonances for specific vibronic transitions are much broader with an incoherent pump, and the overall efficiency of reaching a particular vibrational state is reduced in resonance.
Overall, the efficiency of the two-cavity mode setup for selectively populating vibration states decreases slightly when moving from a molecule to two molecules.

\section{Conclusions}
In summary, we have explored how one or two cavity modes in the \gls{esc} regime can be utilized to selectively populate the excited vibrational states of a molecular system.
This approach uses the decay of cavity photons as a highly effective channel to populate specific vibrational states in combination with a \gls{stirap}-like coupling scheme.
Tuning the frequency of the dumping cavity mode to vibronic transitions between ground and electronically excited states allows for a selective transfer of population into individual vibrational states.
The molecular system is pumped by coherently or incoherently driving the second cavity mode.

To demonstrate this approach, we have used a diatomic model system with two electronic states and an extended version of the \gls{jc} Hamiltonian.
As a reference, we have shown how a molecule behaves when pumped directly by a continuous wave laser field.
This scheme yielded the best result and allowed for $>95$\% to be transferred to
the desired target state.
However, this scheme requires direct access of the
driving laser to the molecule and may not be feasible in microcavities.
When one of the cavity modes is coherently pumped with a laser field,
the efficiency is slightly reduced.
Nevertheless, the efficiency of the coherent pumping scheme improves with increases again for an increased cavity coupling strength.
Even for a detuned dumping cavity mode, vibrational heating is then suppressed more efficiently compared to the direct laser drive.
Thus, this scheme performs best in the \gls{esc} regime and allows a selective pumping of the \gls{stirap}-like pathway.
The electronically excited state is effectively not populated, and the system is protected from vibrational heating due to spontaneous emission.
In contrast, the incoherent pumping scheme still worked well at the boundary between \gls{esc} and weak coupling, but performed worse when the coupling strength increased well into \gls{esc}.
The lack of coherence in the system enables the pumping to occupy states with electronically excited state contribution, leading to vibrational heating.

Extending the multilevel model beyond a single molecule to describe two identical molecules, the two-cavity mode setup efficiently excites both molecules to the same vibrational state.
This process can be rationalized as two consecutive pump-dump cycles.
Compared with the single-molecule setup, the efficiency is slightly reduced.
However, in particular, for the coherently driven cavity mode, the scheme remains
highly selective and efficiently suppresses vibrational heating.

In summary, we have shown that coupling molecules to cavity modes under \gls{esc} can achieve selective population transfer to excited vibrational states when the coupled system is driven.
Although we applied this concept only to a simplified multilevel model, our findings may indicate a possible route to steer and control chemical reactions on an electronic ground state surface.

\begin{acknowledgments}
This project has received funding from the European Research Council (ERC) under the European Union's Horizon 2020 research and innovation program (grant agreement no. 852286). Support from the Swedish Research Council (Grant No. VR 2024-04299) is acknowledged.
\end{acknowledgments}

\section*{Supplementary Material}
See the supplemental material for details on the molecular model and an analysis of the formed polaritonic states using the two-cavity mode setup, as well as additional plots of the propagation of the molecular cavity systems.

\section*{Author Declaration Section}
\subsection*{Conflict of Interest Statement }
The authors have no conflict of interest to disclose.

\subsection*{Author Contributions}
\textbf{Lucas Borges}: Data curation (lead); Formal analysis (equal); Methodology (equal); Visualization (lead); Writing – original draft (equal). \textbf{Thomas Schnappinger}: Conceptualization (equal); Formal analysis (equal); Methodology (equal); Investigation (equal); Writing – original draft (equal), Writing – review \& editing (equal). \textbf{Markus Kowalewski}: Conceptualization (equal); Formal analysis (equal); Funding
acquisition (lead); Methodology (equal);
Project administration (lead); Supervision (lead); Writing – original draft (equal); Writing – review \& editing (equal).

\section*{Data Availability}
The data supporting the findings of this study are available from the corresponding author upon reasonable request.

\end{document}


\title{Supporting Information: \\Selective excitation of molecular vibrations via a two-mode cavity Raman scheme}

\author{Lucas Borges}
\affiliation{Department of Physics, Stockholm University, AlbaNova University Center, SE-106 91 Stockholm, Sweden}

\author{Thomas Schnappinger}
\email{thomas.schnappinger@fysik.su.se}
\affiliation{Department of Physics, Stockholm University, AlbaNova University Center, SE-106 91 Stockholm, Sweden}

\author{Markus Kowalewski}
\email{markus.kowalewski@fysik.su.se}
\affiliation{Department of Physics, Stockholm University, AlbaNova University Center, SE-106 91 Stockholm, Sweden}

\maketitle

\tableofcontents
\clearpage

\section{Details on the molecular model}

\begin{figure}
    \centering
    \includegraphics[width=0.75\textwidth]{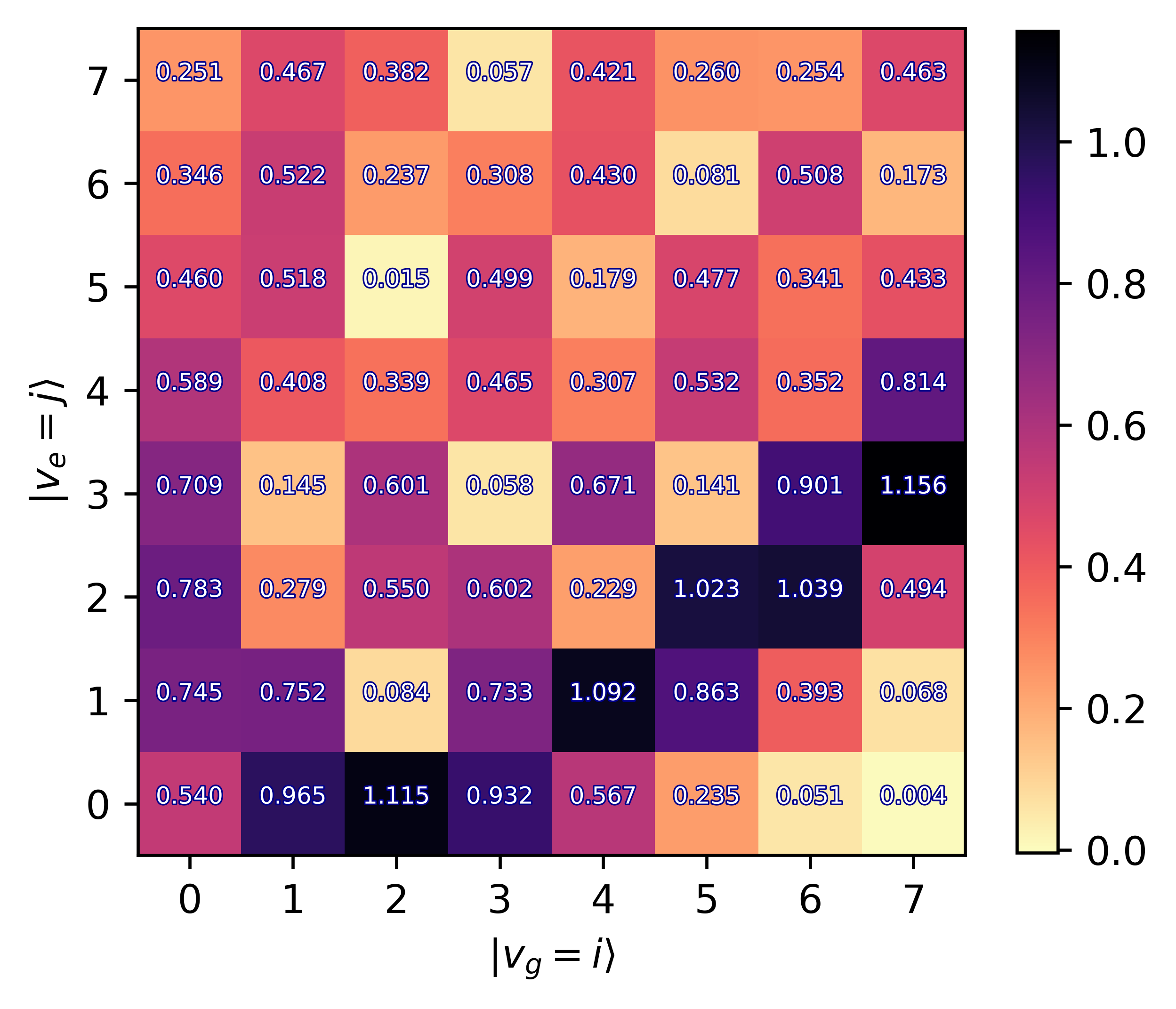}
    \caption{Franck Condon factors for the first eight vibrational states of the first two electronic states  of \mgh}
    \label{fig:FranckCondon_matrix}
\end{figure}

The coupling between the cavity mode $\lambda$ and the vibronic states $\ke{v_g=j}$ and $\ke{v_e=i}$ of the molecules is governed by the coupling strength $g_{ij}^\lambda$ (Eq.~3 in the manuscript).
The coupling strength depends on the Franck-Condon factors $
\mu_{ij} = \expcv{\chi^g_i(R)}{\hat\mu_{eg}(R)}{\chi^e_j(R)}_R$ with the transition dipole moment operator $\hat\mu_{eg}(R)$, the vibrational eigenfunctions $\chi^g_i$ and $\chi^e_j$ of the two electronic states, $e$ and $g$ and the nuclear coordinates $R$.
The underlying ab initio \glspl{pes} of the \mgh molecule were previously used to study the effect of \gls{esc} on molecular systems, and details of the calculations can be found in references~\cite{Davidsson2020-qb,Davidsson2020-bs,Borges2024-pn}.
Fig.~\ref{fig:FranckCondon_matrix} shows the Franck-Condon factors for the first eight vibrational states of both electronic states of \mgh.
\clearpage
The energy differences between the vibronic states formed by the first four vibrational states of $e$ and $g$ are given in Table~\ref{tab:wegs_table}.
\begin{table}[h]
    \centering
    \begin{tabular}{l|ccccccc}
        $\Delta\omega_{ij}$ & $v_g=0$  & $v_g=1$ & $v_g=2$ & $v_g=3$ & $v_g=4$ & $v_g=5$ & $v_g=6$\\
        \hline
        $v_e=0$ & 4.3218 & 4.1209 & 3.9282 & 3.7441 & 3.5690 & 3.4033 & 3.2469 \\
        $v_e=1$ & 4.4579 & 4.2569 & 4.0642 & 3.8801 & 3.7050 & 3.5393 & 3.3830 \\
        $v_e=2$ & 4.5921 & 4.3911 & 4.1985 & 4.0144 & 3.8393 & 3.6735 & 3.5172 \\
        $v_e=3$ & 4.7245 & 4.5235 & 4.3308 & 4.1468 & 3.9717 & 3.8059 & 3.6496 \\
        $v_e=4$ & 4.8541 & 4.6531 & 4.4604 & 4.2764 & 4.1013 & 3.9355 & 3.7792 \\
    \end{tabular}
    \caption{Energy difference between vibronic states formed by the first $n_g=7$ and $n_e=5$ vibrational states of $\ke{g}$ and $\ke{e}$ in eV, respectively.}
    \label{tab:wegs_table}
\end{table}

\section{Analysis of the polaritonic states} \label{sec:SI_Eigenstates}

\begin{figure}
    \centering
    \includegraphics[width=10cm]{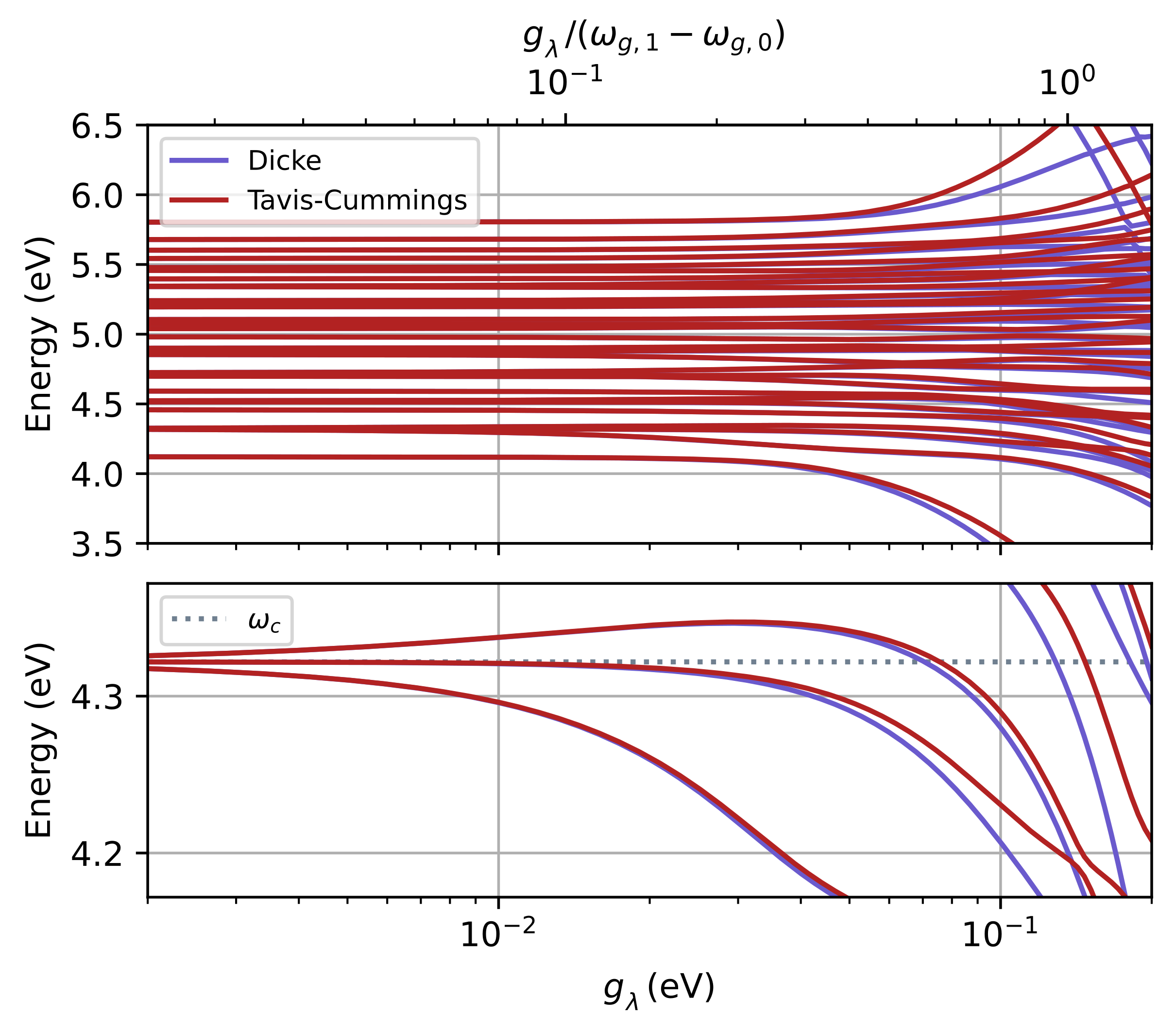}
    \caption{Eigenvalues for the Tavis-Cummings and Dicke models (with and without the rotating wave approximation) as a function of the cavity coupling strength $g_{\lambda} = g_{00}^p=g_{00}^d$.
    The molecular model includes ten vibrational states for both electronic states and is resonantly coupled to the two cavity modes $p$ and $d$ ($\omega_p = \Delta\omega_{00} = 4.3218$ eV and $\omega_d = \Delta\omega_{10} = 4.1209$ eV).
    The Dicke model includes all the states up to the fourth excitation manifold (i.e., four photons).
    The lower panel zooms on the eigenvalues that involve the states $\ke{v_g=0;1_p0_d}, \ke{v_g=1;0_p1_d}$ and $\ke{v_e=0;0_p0_d}$.}
    \label{fig:RWA_eig_check}
\end{figure}

To validate the use of the rotating wave approximation, we compare the eigenenegy spectrum of the molecular model resonatly coupled to the two cavity modes $p$ and $d$ ($\omega_p = \Delta\omega_{00} = 4.3218$ eV and $\omega_d = \Delta\omega_{10} = 4.1209$ eV) calculated using the Tavis-Cumming model~\cite{Tavis1967-op} with the one calculated using the Dicke model~\cite{Dicke1954-sq}.
The molecular model used includes ten vibrational states per electronic state.
The two Hamiltonians differ by the presence of counter-rotating terms in the interaction between the molecular model and the cavity modes.
Consequently, the different excitation manifolds of the Dicke model can no longer be diagonalized independently.
To achieve proper diagonalization, we truncated the Dicke Hamiltonian to include up to the fourth excitation manifold (i.e., four photons). In contrast, for the \gls{jc} Hamiltonian, we only included the first excitation manifold (i.e., one photon).
Both eigenenegy spectra are shown in Figure~\ref{fig:RWA_eig_check} as a function of the cavity coupling strength $g_{\lambda} = g_{00}^p=g_{00}^d$.

For coupling strengths greater than 20\,meV, the eigenstates formed by the bare states $\ke{v_g=0;1_p0_d}$, $\ke{v_g=1;0_p1_d}$, and $\ke{v_e=0;0_p0_d}$ are shifted to lower energies for both the Tavis-Cummings model and the Dicke model.
This shift is due to the interaction with higher-lying eigenstates.
Significant differences in the eigenvalue spectra obtained with and without the rotating wave approximation become evident when the cavity coupling strength exceeds 50\,meV.
For the coupling strength used in this work ($g_{ij}^p=g_{ij}^d \leq 5.00$\,meV), both eigenvalue spectra are nearly identical, and consequently, the rotating wave approximation can be used safely.
Within this range of coupling strengths, the coupling between the relevant vibronic states and cavity modes can be approximated as a three-level system.

\begin{figure}[h]
    \centering
    \includegraphics[width=12cm]{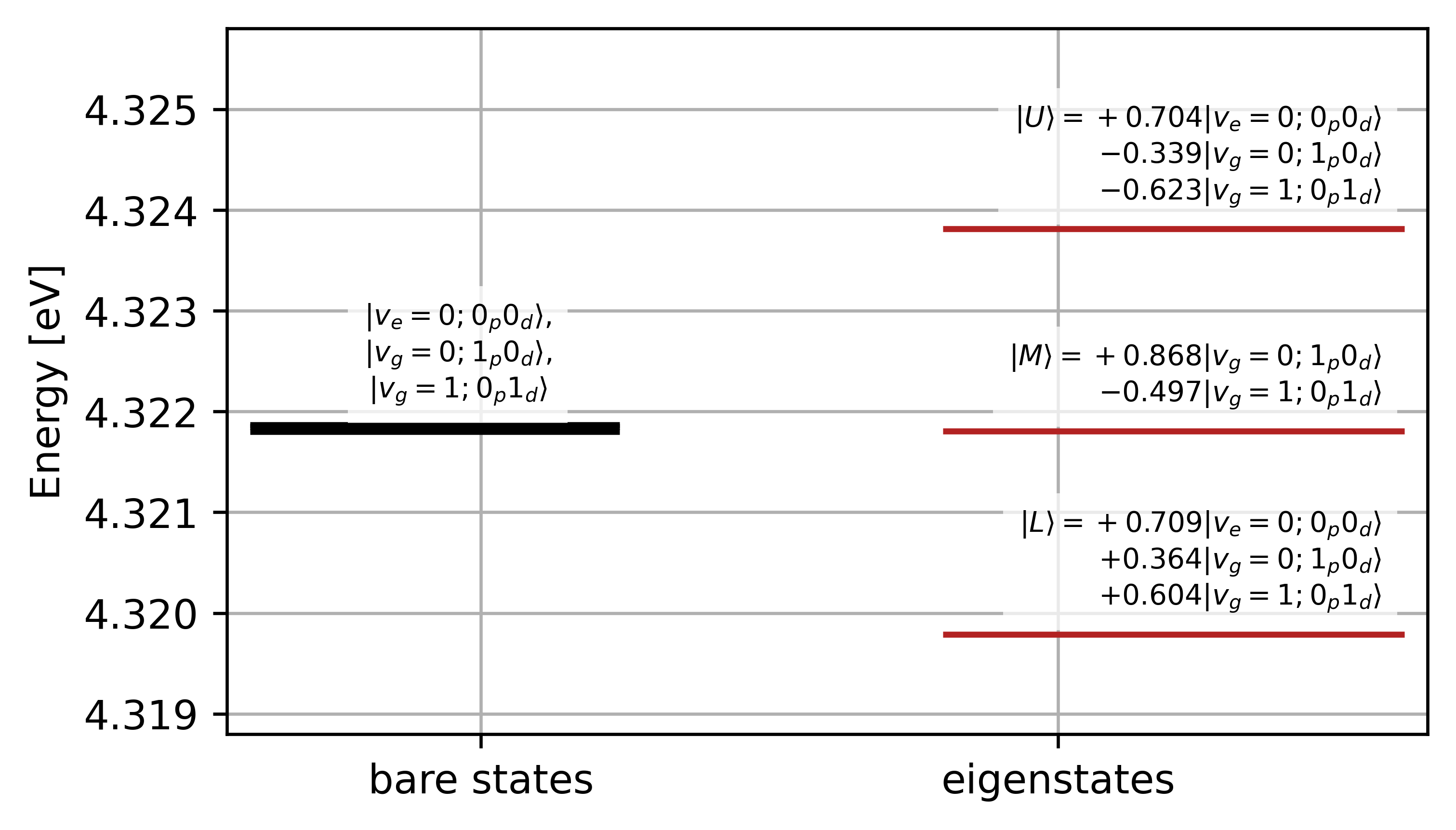}
    \caption{Eigenstates formation for the case of a single molecule coupled to two modes, with the corresponding bare states character $c_j=\bk{\psi_{eig}}{\phi_j}$ written above the bars.
    The bare states shown in the left panel $\ke{v_e=0;0_p0_d}$, $\ke{v_g=1;0_p1_d}$ and $\ke{v_g=0;1_p0_d}$ are at resonance with $\omega_p=\Delta\omega_{00}$ and $ \omega_d =\Delta\omega_{10}$.
    The coupling strengths used are $g_{00}^p=g_{00}^d\sqrt{\omega_p/\omega_d}=1.00$ meV.}
    \label{fig:3LS_eigenstates}
\end{figure}

In the following, we analyze the structure of the first three eigenstates of the first excitation manifold formed by the three bare states $\ke{v_e=0;0_p0_d}$, $\ke{v_g=1;0_p1_d}$, and $\ke{v_g=0;1_p0_d}$.
The energy diagram of the three eigenstates for a coupling strength of $g_{00}^p=g_{00}^d = 1.00$\,meV is shown in Fig.~\ref{fig:3LS_eigenstates}.
The three polaritonic eigenstates closely resemble the eigenstates of a three-level system used in \gls{stirap}.
Both the upper eigenstate ($\ke{U}$) and the lower eigenstate ($\ke{L}$) have contributions from the excited state $\ke{v_e=0;0_p0_d}$, while the middle eigenstate ($\ke{M}$) does not include this contribution.
For coupling strengths $g_{00}^p=g_{00}^d<10$\,meV, the contributions of other bare states are negligible, and the energy difference between the eigenstates $\ke{U}$ and $\ke{L}$ is given by $2\sqrt{(g_{00}^p)^2+(g_{10}^d)^2} \approx 4g_{00}^p$.

\begin{figure}[h]
    \centering
    \includegraphics[width=\textwidth]{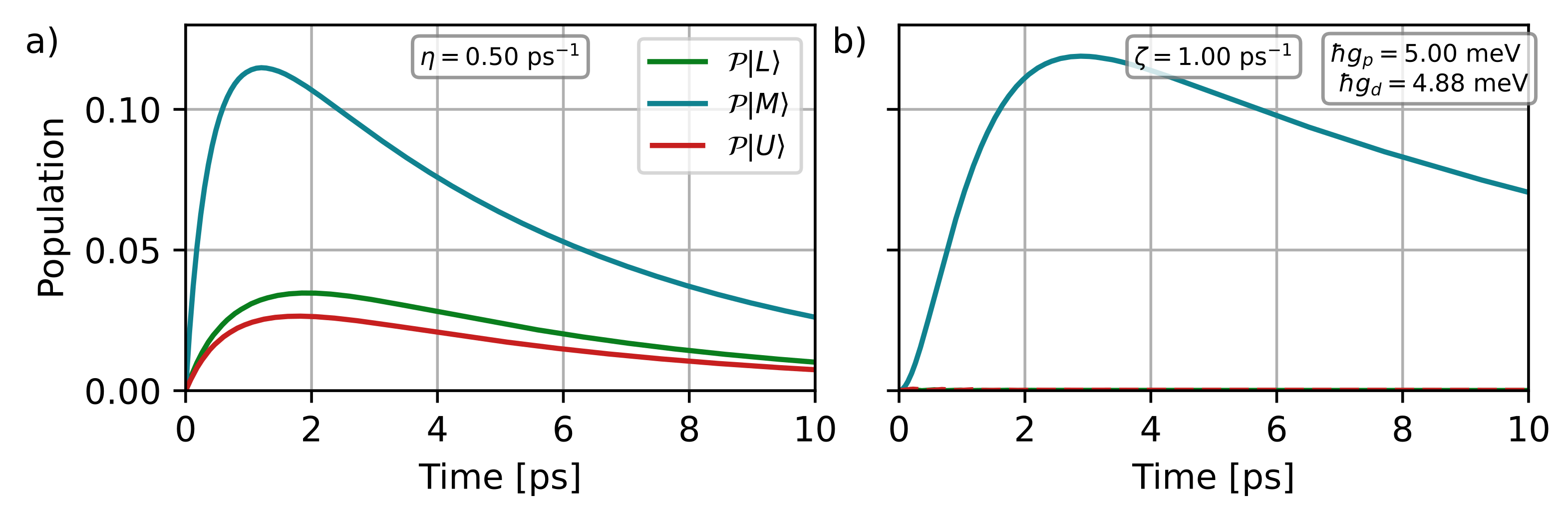}
    \caption{Populations of the three polaritonic eigenstates $\ke{U}$, $\ke{M}$, and $\ke{L}$ for a single molecule coupled to two modes with different pumping schemes.
    The initial state $\ke{v_g=0,0_p0_d}$ is pumped by (a) an incoherent pump with $\eta = 0.50$ ps$^{-1}$ and (b) a coherent pump with $\eta = 1.00$ ps$^{-1}$.
    The cavity frequencies are $\omega_p=\Delta\omega_{00}$ and $ \omega_d =\Delta\omega_{10}$.
    The coupling strength is set to $g^p_{00}=g^d_{00}\sqrt{\omega_p/\omega_d}=5.00$ meV. }
    \label{fig:eigen_projection}
\end{figure}

Figure~\ref{fig:eigen_projection} shows the projection of the initial population dynamics on the three polaritonic eigenstates $\ke{U}$, $\ke{M}$ and $\ke{L}$ for a single molecule coupled to two cavity modes and a coupling strength of $g^p_{00}=g^d_{00}\sqrt{\omega_p/\omega_d}=5.00$ meV.
In the case of incoherent pumping of the cavity mode $p$, shown in Figure~\ref{fig:eigen_projection}(a), all three polaritonic eigenstates, $\ke{U}$, $\ke{M}$, and $\ke{L}$, are populated, with $\ke{M}$ being the most populated.
In contrast, the coherent pumping shown in Figure~\ref{fig:eigen_projection}(b) only selectively populates $\ke{M}$.
As the state $\ke{M}$ has no contribution from the electronically excited state $\ke{e}$, it is responsible for efficient population transfer between the two vibrational
states.
The other two states, $\ke{U}$ and $\ke{L}$, have contributions from the excited state $\ke{e} $, and consequently their presence leads to vibrational heating, resulting in a reduced overall selectivity compared to the coherent pumping scheme.

\clearpage

\section{Supplemental plots for the dynamics}

\subsection{Performance of the molecular model}

\begin{figure}
    \centering
    \includegraphics[width=10cm]{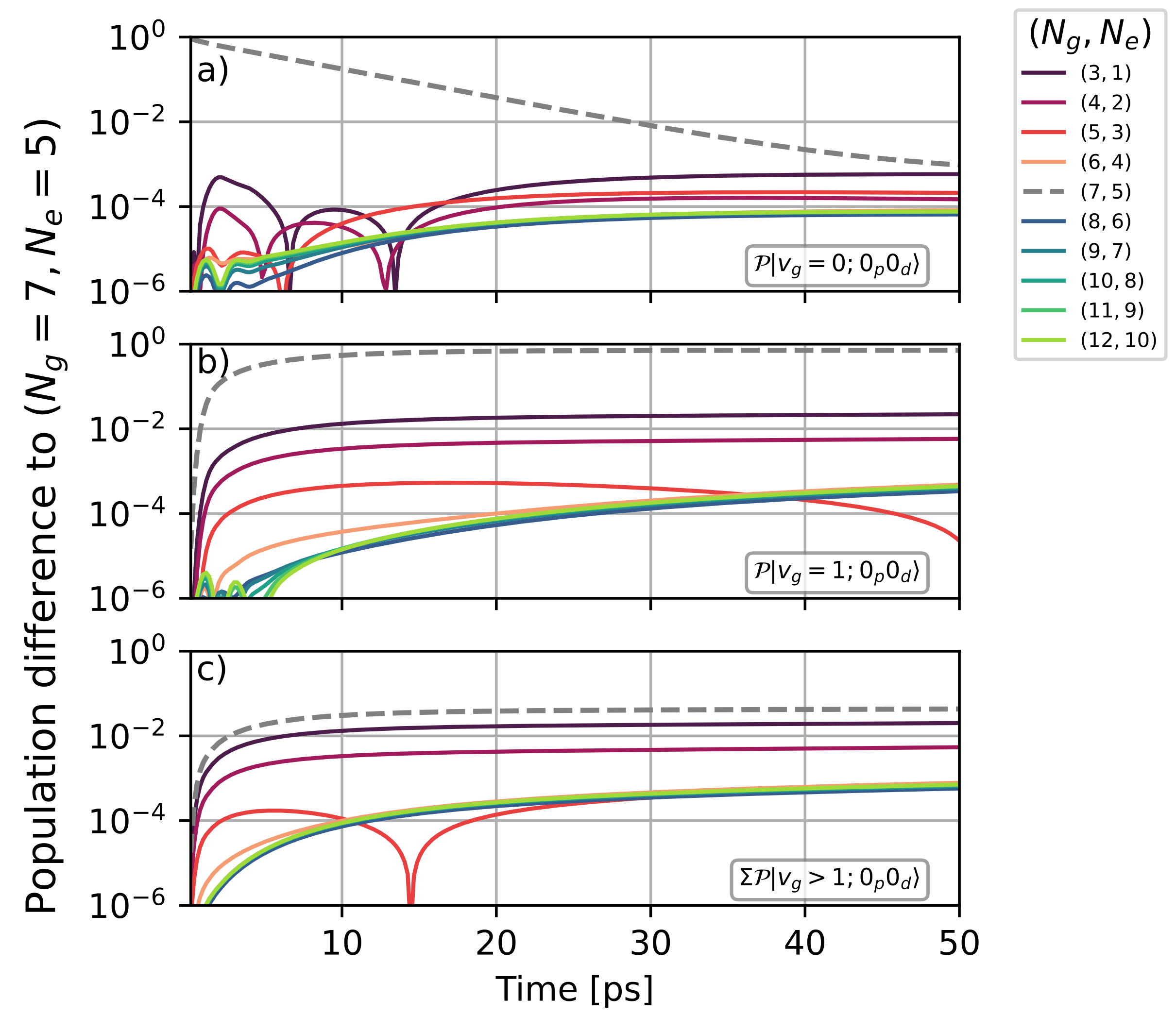}
    \caption{Time evolution of a single molecule coupled to two cavity modes $p$ and $d$ which are tuned to $\omega_p=\Delta\omega_{00}$, $\omega_d=\Delta\omega_{10}$  and $\kappa_p = \kappa_d =2.00\,\text{ps}^{-1}$.
    Difference in the time-dependent populations of states (a) $\ke{v_g=0;0_p0_d}$, (b) $\ke{v_g=1;0_p0_d}$, and (c) the sum $\Sigma\mathcal{P}\ke{v_g>1;0_p0_d}$ for different values of $n_g$ and $n_e$, relative to the result with $n_g=7$ and $n_e=5$.
    The cavity mode $p$ is incoherently pumped with $\eta=0.50\,\text{ps}^{-1}$.
    The spontaneous decay rate $\Gamma_{00}$ is 1/40 ps$^{-1}$.
    The dashed gray lines represent the corresponding populations of the propagation with $n_g=7$ and $n_e=5$.
    }
    \label{fig:convergence_N1}
\end{figure}

The multilevel model used to approximate a single \mgh molecule in this work is truncated to include seven vibrational states in the electronic ground state ($n_g=7$) and five vibrational states in the electronic excited state ($n_e=5$).
To justify this truncation, simulations of the molecular model coupled to a two-mode cavity are performed with varying numbers of vibrational states included in the Hamiltonian.
Figure~\ref{fig:convergence_N1} shows the difference in the time-dependent populations of states $\ke{v_g=0;0_p0_d}$, $\ke{v_g=1;0_p0_d}$, and the sum $\Sigma\mathcal{P}\ke{v_g>1;0_p0_d}$ for different values of $n_g$ and $n_e$, relative to the result with $n_g=7$ and $n_e=5$.
The dashed gray curves in Fig.~\ref{fig:convergence_N1} represent the population of the propagation with $n_g=7$ and $n_e=5$.
The difference in populations for propagations with a larger Hamiltonian, which includes more vibrational states, is less than $10^{-3}$, which is significantly smaller than the population changes discussed.
Consequently, increasing the number of vibrational states in the Hamiltonian beyond $n_g=7$ and $n_e=5$ results only in slight variations that do not justify the additional computational time required to calculate the propagations.

\begin{figure}[b]
    \centering
    \includegraphics[width=\textwidth]{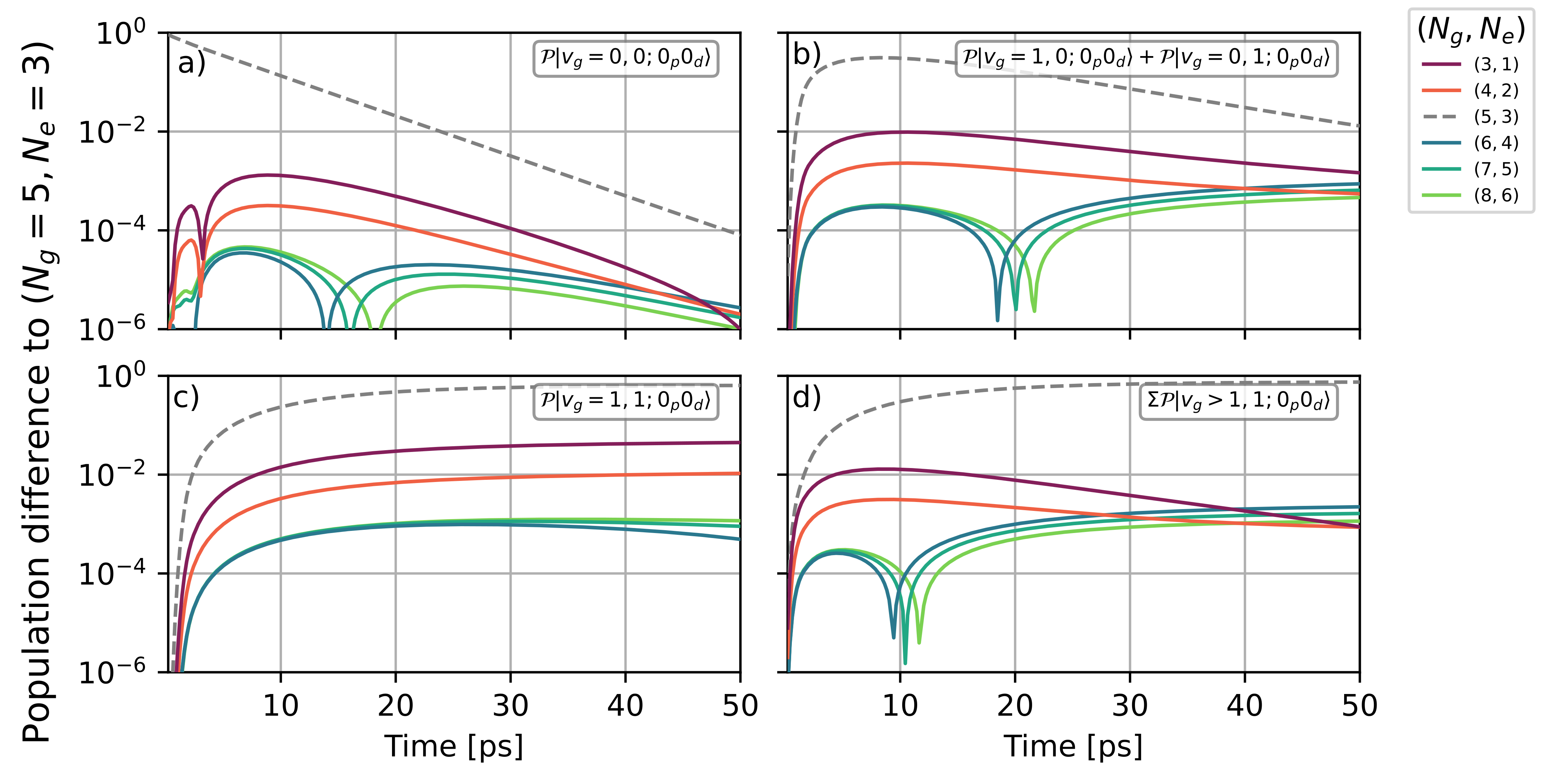}
    \caption{
    Temporal evolution of the populations of two molecules coupled to two cavity modes $p$ and $d$ which are tuned to $\omega_p=\Delta\omega_{00}$, $\omega_d=\Delta\omega_{10}$  and $\kappa_p = \kappa_d =2.00\,\text{ps}^{-1}$.
Difference in the time-dependent populations of states (a) $\mathcal{P}\ke{v_g=0,0;0_p0_d}$, (b) $\mathcal{P}\ke{v_g=1,0;0_p0_d}+\mathcal{P}\ke{v_g=0,1;0_p0_d}$, (c) $\mathcal{P}\ke{v_g=1,1;0_p0_d}$, and (d) the sum $\Sigma\mathcal{P}\ke{v_g>1;0_p0_d}$ for different values of $n_g$ and $n_e$, relative to the result with $n_g=5$ and $n_e=3$.
    The cavity mode $p$ is incoherently pumped with $\eta=0.50\,\text{ps}^{-1}$.
    The spontaneous decay rate $\Gamma_{00}$ is 1/40 ps$^{-1}$.
        The dashed gray lines represent the corresponding populations of the propagation with $n_g=5$ and $n_e=3$.}
    \label{fig:convergence_N2}
\end{figure}

For the case of two identical molecules, we truncated the multilevel model for each of the molecules to include only five vibrational states in the electronic ground state ($n_g=5$) and five vibrational states in the electronic excited state ($n_e=3$).
The difference in the time-dependent populations of the relevant bare states is shown in Figure~\ref{fig:convergence_N2} for different values of $n_g$ and $n_e$ per molecule, relative to the result with $n_g=5$ and $n_e=3$.
Similarly to the single molecule case, the propagations that include additional vibrational states result in minor variations in the populations, on the order of $ 10^{-3}$.
Consequently, truncating to $n_g=5$ and $n_e=3$ per molecule is sufficient to discuss the relevant changes in population during the simulation.

\subsection{Photostationary state of the empty cavity}

\begin{figure}[h]
    \centering
    \includegraphics[width=10cm]{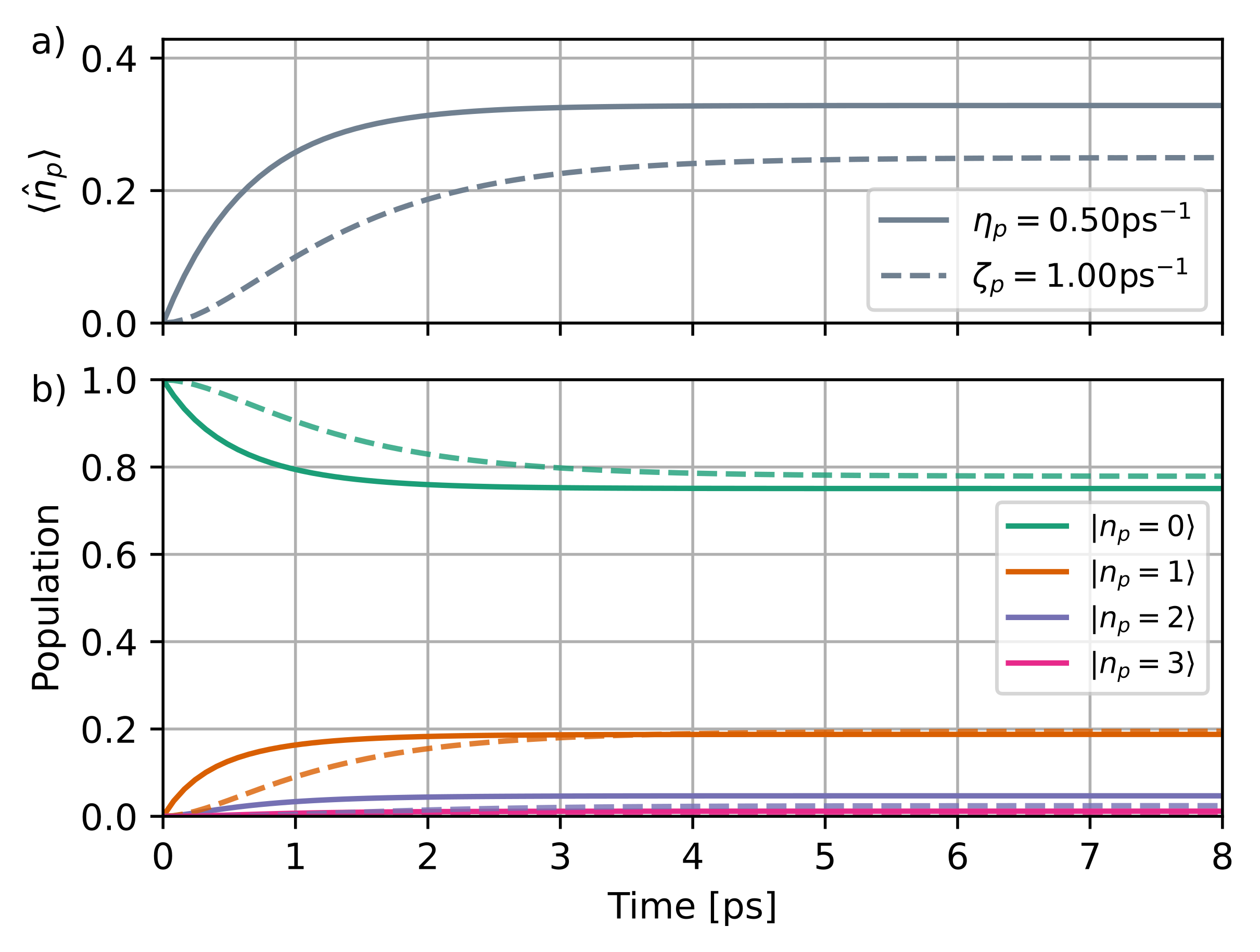}
    \caption{Temporal evolution of the empty two-mode cavity subjected to photon pumping and decay. (a) Photon number expectation value and (b) Fock states populations.
    The cavity mode $p$ is pumped incoherently and coherently with the rates $\eta=0.50\text{ps}^{-1}$ and $\zeta=1.00\text{ps}^{-1}$, respectively.
    The photon decay rate $\kappa$ is $ 2.00\,\text{ps}^{-1}$.}
    \label{fig:empty_cavity}
\end{figure}

The time evolution of the empty cavity is shown in Fig.~\ref{fig:empty_cavity}.
The cavity mode $p$ is driven by incoherent pumping ($\eta\create_p$) and coherent pumping ($\zeta(\create_p+\destroy_p)\cos(\Delta\omega_{00}t)$), represented by continuous and dashed curves, respectively.
The expectation value of the photon number differs between the two pumping schemes (Fig.~\ref{fig:empty_cavity}a) for the chosen pump rates $\eta = \zeta/2 = 0.50\text{ps}^{-1}$.
However, the population of the Fock states, the eigenstates of the photon number operator, is very similar (Fig.~\ref{fig:empty_cavity}b).

\clearpage
\subsection{Additional propagations for the single molecule case}

\begin{figure}[h]
    \centering
    \includegraphics[width=\textwidth]{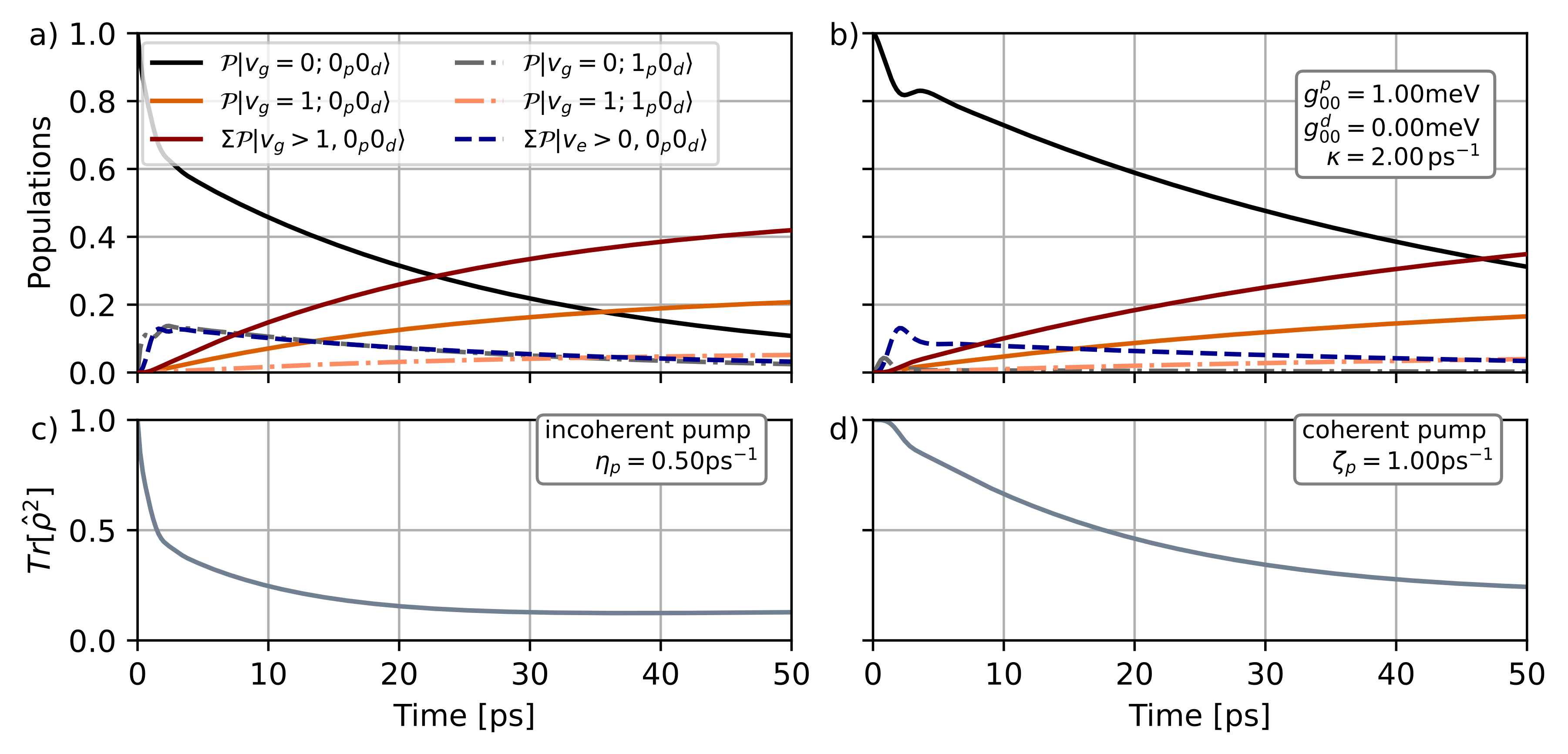}
    \caption{
    Time evolution of a single molecule pumped by a single cavity mode $p$. Temporal evolution of the populations $\mathcal{P}\ke{\phi,n_d}$ (a)-(b) and the purity of the density matrix $\text{Tr}[\hat{\rho}^2]$ (c)-(d).
    The cavity mode $p$ is resonant with the $\ke{v_g\!=\!0}\rightarrow\ke{v_e\!=\!0}$ transition and the cavity decay $\kappa_p$ is 2.00\,ps$^{-1}$.
    For (a) and (c), the cavity mode $p$ is incoherently pumped according to $\eta=0.50\,\text{ps}^{-1}$. For (b) and (d), the cavity mode $p$ is coherently pumped according to Eq.~7 in the manuscript with $\zeta=1.00\,\text{ps}^{-1}$.
    The spontaneous decay rate $\Gamma_{00}$ is 1/40 ps$^{-1}$.}
    \label{fig:SingleMode_comparison_pumps}
\end{figure}

Figure~\ref{fig:SingleMode_comparison_pumps} shows the time evolution of a molecule coupled to a single-mode cavity, where the cavity mode $p$ is resonant to the 0-0 transition and pumped by the incoherent pump scheme and the coherent pump scheme.

The spontaneous decay leads to the populations of multiple vibrational states in the ground states, represented by the dark red curves.
The observed low purity of the density matrix is an indication of a vibrational hot state.
\clearpage

\begin{figure}[h]
    \centering
    \includegraphics[width=\textwidth]{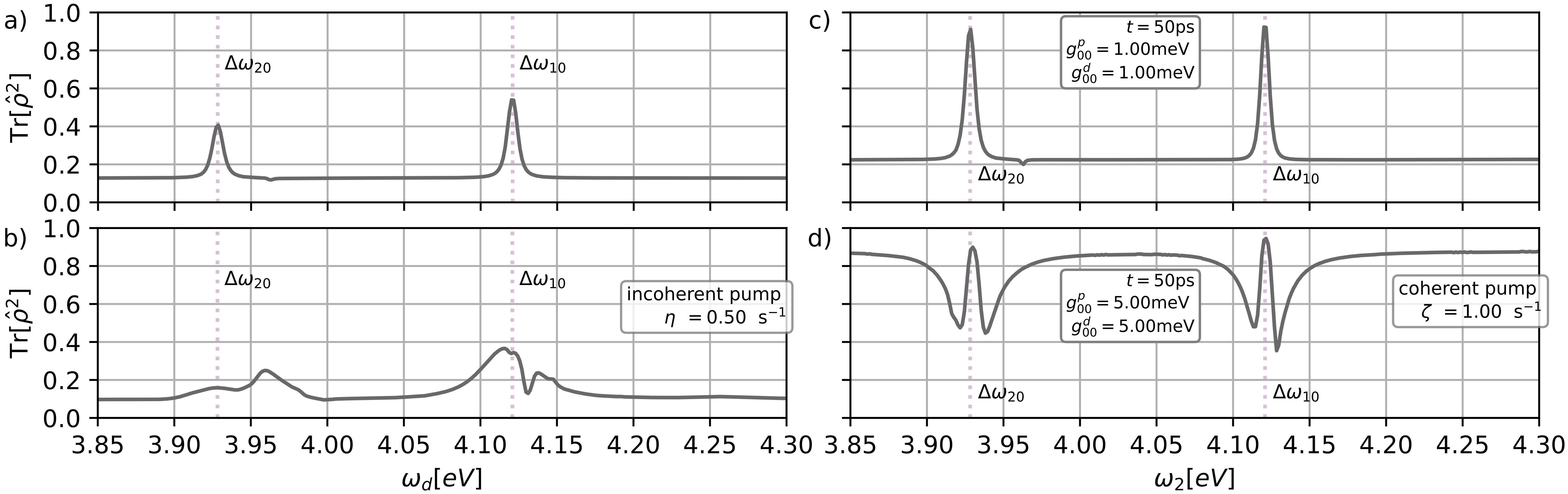}
    \caption{
    Frequency scan of cavity mode $d$ for a single molecule driven by cavity mode $p$.
    Purity of the density matrix $\text{Tr}[\hat{\rho}^2]$ as function of the cavity frequency $\omega_d$ after $t=50$\,ps for a single molecule coupled the two cavity modes $p$ and $d$ with $\omega_p=\Delta\omega_{00}$, and $\kappa_p = \kappa_d =2.00\,\text{ps}^{-1}$. For (a)-(b), the cavity mode $p$ is incoherently pumped with $\eta=0.50\,\text{ps}^{-1}$. For (c)-(d), the cavity mode $p$ is coherently pumped with $\zeta=1.00\,\text{ps}^{-1}$.
    The coupling strengths of both modes are given by $g_d = \sqrt{\omega_d/\omega_p}g_{p}$, where $g_{p}$ is set to (a)-(c) $1.00$\,meV and (b)-(d) $5.00$\,meV.
    The spontaneous decay rate $\Gamma_{00}$ is $1/40\,\text{ps}^{-1}$, and the vertical dotted lines are indicating resonaces $\Delta\omega_{i0} = \omega_{e,0}-\omega_{g,i}$ with $i=1,2$.
    }
    \label{fig:N1_scanw2_purity_1meV}
\end{figure}

To further evaluate the selectivity of the proposed pump-dump scheme using two cavity modes, we scan the frequency of the cavity mode $d$ while keeping the cavity mode $p$ resonant with the $\ke{v_g\!=\!0}\rightarrow\ke{v_e\!=\!0}$ transition.
Figure~\ref{fig:N1_scanw2_purity_1meV} shows the purity of the density matrix $\text{Tr}[\hat{\rho}^2]$ as a function of $\omega_d$ for different cavity coupling strengths and both pumping scenarios.

\subsection{Scan of the coupling strength}

\begin{figure}[h]
    \centering
    \includegraphics[width=\textwidth]{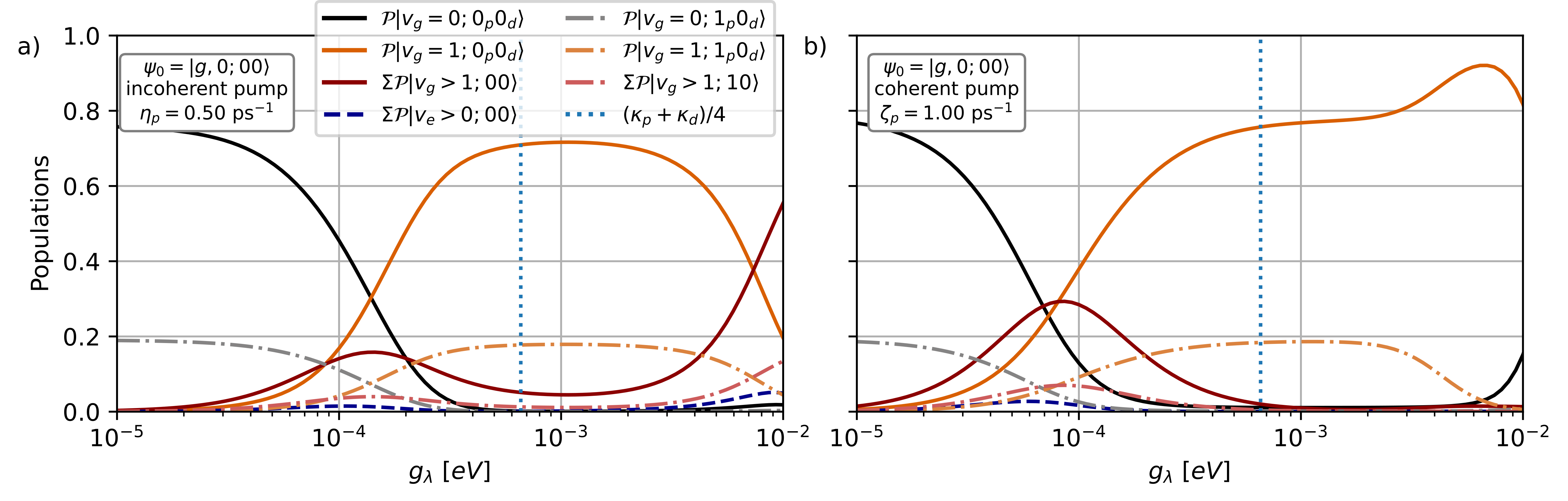}
    \caption{
 Scan of cavity coupling strengths for a single molecule coupled to two cavity modes $p$ and $d$.
    Populations $\mathcal{P}\ke{\phi,n_p,n_d}$ as function of $g_\lambda=g_{00}^d=g_{00}^p$ after $t=50$\,ps for $\omega_p=\Delta\omega_{00}$, and $\kappa_p = \kappa_d =2.00\,\text{ps}^{-1}$. For (a) the cavity mode $p$ is incoherently pumped with $\eta=0.50\,\text{ps}^{-1}$. For (b) the cavity mode $p$ is coherently pumped with $\zeta=1.00\,\text{ps}^{-1}$.
    The relation between the coupling strengths of both modes is given by $g_d = \sqrt{\omega_d/\omega_p} g_p$. The spontaneous decay rate $\Gamma_{00}$ is 1/40\,ps$^{-1}$, and the vertical blue dotted lines indicate the cavity decay rates $(\kappa_p+\kappa_)/4$.}
    \label{fig:N1_scan_g}
\end{figure}

The population after 50\,ps of a single molecule coupled to two cavity modes for incoherent and coherent pumping as a function of cavity coupling strengths $g_\lambda=g_{00}^d=g_{00}^p$ is shown in Fig.~\ref{fig:N1_scan_g}.

On the right-hand side of the blue dotted lines in Figs.~\ref{fig:N1_scan_g}~a) and b), the coupled molecular cavity system is in the strong coupling regime with $g_{00}^p > \kappa_p/4$.
For both incoherent and coherent pumping, the population transfer to the target state $\ke{v_g=1; 0_p0_d}$ occurs already at coupling strengths below $\kappa_p/4$.
Interestingly, the onset of population transfer into the target state is observed for smaller coupling strengths for coherent pumping than for incoherent pumping.
However, this transfer is accompanied by a significant occupation of higher vibrational states of $\ke{g}$ (depicted by the dark red curves).
This unwanted heating is strongest around $g_\lambda=10^{-4}$\,eV and more pronounced in the case of coherent pumping.
When transitioning to the strong coupling regime, the population in higher vibrational states decreases in both pumping scenarios, while the population in the target state increases.
For coherent pumping, the heating is completely suppressed, whereas for incoherent pumping, some population remains in higher vibrational states and above $g_\lambda=10^{-3}$\,eV, this population increases again.